\documentclass[twocolumn,times,tighten]{aastex62}

\graphicspath{{./}{figures/}}

\submitjournal{ApJ}
\received{June 20, 2018}
\revised{October 5, 2018}
\accepted{October 17, 2018}

\shorttitle{The physical and chemical properties of  $\rho$ oph A}
\shortauthors{Chen \& Hirano}
\begin{document}

\title{The physical and chemical properties of the $\rho$ ophiuchi A dense core}

\correspondingauthor{Yu-Ching Chen}
\email{ycchen@illinois.edu}

\author[0000-0002-9932-1298]{Yu-Ching Chen}
\affiliation{Department of Astronomy, University of Illinois at Urbana-Champaign, 1002 W. Green Street, Urbana, IL 61801, USA}
\affiliation{National Center for Supercomputing Applications, University of Illinois at Urbana-Champaign, 1205 West Clark St., Urbana, IL 61801, USA}

\author{Naomi Hirano}
\affiliation{Academia Sinica Institute of Astronomy and Astrophysics, P.O. Box 23-141, Taipei 106, Taiwan}



\begin{abstract}

The physical and chemical properties of the $\rho$ Ophiuchi A core were studied using 1.3 mm continuum and molecular lines such as C$^{18}$O, C$^{17}$O, CH$_3$OH and H$_2$CO observed with the Submillimeter Array (SMA). 
The continuum and C$^{18}$O data were combined with the single-dish data obtained with the IRAM 30m telescope and the James Clerk Maxwell Telescope (JCMT), respectively. The combined 1.3 mm continuum map reveals three major sources, SM1, SM1N and VLA1623 embedded in the extended emission running along the north-south direction, and two additional compact condensations in the continuum ridge connecting SM1 and VLA1623.
The spatial distribution of the C$^{18}$O emission is significantly different from that of the continuum emission; the C$^{18}$O emission is enhanced at the eastern and western edges of the continuum ridge, with its peak brightness temperature of 40--50 K.
This supports the picture that the $\rho$-Oph A core is heated externally from the nearby stars Oph S1 and HD147889.
In contrast, the C$^{18}$O intensity is lower than 15--20 K at the center of the ridge where the continuum emission is bright.
The C$^{18}$O abundance decreases inside the ridge, and shows anti-correlation with the N$_2$H$^+$ abundance. However, both C$^{18}$O and N$_2$H$^+$ show strong depletion at the Class 0 protostar VLA1623, implying that the dense gas surrounding VLA1623 is colder than the freeze-out temperature of N$_2$. The blue- and red-shifted components of CH$_3$OH and H$_2$CO lines are seen at SM1, suggesting outflow activity of embedded source in SM1, although the spatial distributions do not show clear bipolarity.

\end{abstract}



\section{Introduction} \label{sec:intro}

It is well established that the formation of low-mass stars occurs from the collapse of gravitationally-bound molecular cloud cores \citep[e.g.][]{Shu1987}.  Cores before the onset of star formation are called ^^ ^^ pre-protostellar cores" or ^^ ^^ pre-stellar cores".
Physical and chemical properties of prestellar cores on the verge of star formation are particularly interesting because they provide us the clues to understand the initial conditions of star formation. 
%These so-called "pre-protostellar" clouds are thought to be precursors of Class 0 objects, the youngest accreting protostars, which were already found and studied in several regions\citep{1998A&A...336..150M,2015ApJ...805..125T}. However, the understanding of the pre-protostellar cloud remains poorly known so far. By probing the chemical abundance and gas dynamics via molecular line emission, the researchers could analyze the process of this stage and then supplement the current star formation theory.

The $\rho$ Ophiuchus cloud complex at a distance of 137.3$\pm$1.2 pc \citep{Ortiz-Leon2017} is one of the closest molecular clouds with ongoing low-mass star formation. The $\rho$ Ophiuchus main cloud L1688 consists of several regions labeled from A to F \citep[e.g.][]{Loren1990}; among them, the $\rho$-Oph A core is the most prominent core in the millimeter and submillimeter continuum emission \citep[e.g.][]{Motte1998,Wilson1999,Johnstone2000}.
%\citep{1984A&A...141..127Z,1998A&A...336..150M}. 
The continuum images show that the $\rho$-Oph A  is a curved ridge consisting of a chain of condensations labeled SM1, SM1N, SM2 and  VLA1623. 
Except VLA1623, which is a widely-studied Class 0 object \citep{Andre1993}, 
the other condensations in the $\rho$-Oph A core have been considered to be starless, because there was no clear signature of star formation.
%the \sout{The}} physical properties \textcolor{red}{and the star formation signature in these \sout{of those tiny}} condensations are still not \textcolor{red}{\sout{clear due the limitation of the resolution} well understood}. 
However, recent high resolution and sensitivity observations with the Acatama Large Millimeter/Sub-Milimeter Array (ALMA) at Band 7 have revealed that SM1 contains a spatially compact continuum emission component with a size of $\sim$40 AU \citep{Friesen2014}.
\citet{Friesen2014} also claimed that SM1 is protostellar because of the X-ray and radio detection toward this source.
Later ALMA observations at Band 3 discovered another compact continuum source between SM1 and VLA1623 \citep{Kirk2017}.
On the other hand, \citet{Nakamura2012} combined the data obtained with the Submillimeter Array (SMA) and single-dish, and found additional small condensations within SM1.
Their results demonstrated the importance of filling the short-spacing data in order to find the small-scale condensations buried in the extended emission.
%Though some high-resolution interferometer like Acatama Large Millimeter/Sub-Milimeter Array (ALMA) could probe individual core, it is hard for it to measure the physical quantities like column density for the whole $\rho$-Oph A cloud due to the lack of short baseline. Thanks to the techniques of combining the single-dish data and interferometer data\citep{2003nro...65}, we are able to obtain the full coverage of baseline and still remain high spatial resolution, which provide us good flux measurement and morphology with high resolution. 

In order to examine the evolutionary stage of the condensations, kinematic information and chemical properties provided by the molecular line observations are helpful.
%In contrast to the continuum observations, high resolution molecular line observations are still missing.
However, most of the interferometric observations have been limited toward the Class 0 protostar VLA1623 \citep[e.g.][]{Murillo2013a,Murillo2013}

In this paper, we present the 1.3 mm continuum and C$^{18}$O (2--1) images obtained with the combination of the SMA and the single-dish data.
Our combined images successfully recovered from large-scale to small-scale ($\sim$5\arcsec) emission from the $\rho$-Oph A region.
We also present the SMA images of the C$^{17}$O, CH$_3$OH, and H$_2$CO from the SM1 region.

%In order to address the question about how those prestellar cores in $\rho$-Oph A evolve and what the physical and chemical condition of those cores are, we combined the single dish data from IRAM 30m telescope and James Clerk Maxwell Telescope(JCMT) 15 telescope with interferometer data from Sub-millimeter Array (SMA). The column density and abundance of C$^{18}$O and N2H$^+$ were calculated and C$^{18}$O molecules appears depleted inside the $\rho$-Oph A continuum ridge. Two new small condensations, VLA1623-N1 and VLA1623-N2 were also discovered in 1.3 mm continuum map. This paper is presented as the following order, observations and data in Section 2, results and analysis in Section 3, Discussions in Section 4 and conclusions in Section 5. 

\section{Observations}
\subsection{SMA data of the $\rho$-Oph A ridge}
The SMA data of the 1.3 mm continuum and the C$^{18}$O (2--1) line were obtained from the SMA data archive. The observations were made on August 5, 2005 in the compact configuration. The primary beam size (FWHM) of the SMA 6-m antennas at 1.3 mm is $~$54\arcsec. The $\rho$-Oph A ridge that is prominent in the continuum map \citep[e.g.][]{Motte1998} was covered with the 4-pointing mosaic (Figure \ref{fig:cont_4point}). The spectral correlator that consists of 24 chunks of 104 MHz bandwidth covers 2 GHz bandwidth in each of the two sidebands separated by 10 GHz. The C$^{18}$O line was observed simultaneously with the CO (2--1) and $^{13}$CO (2--1) lines. The spectral resolution of the chunks with the C$^{18}$O and $^{13}$CO lines is 203 kHz per channel, which corresponds to the velocity resolution of 0.264 km s$^{-1}$. The rest of the channels are covered with the uniform resolution of 812 kHz (1.048 km s$^{-1}$). The absolute flux density scale was determined from the observations of Callisto and Uranus. A pair of nearby quasars, 1626-298 and 1517-243, were used to calibrate the relative amplitude and phase. The bandpass was calibrated by Jupiter and 3c454.3. The visibility data were calibrated using the MIR package in IDL. The continuum data were obtained by averaging the line-free chunks of both sidebands. To improve the signal-to-noise ratio, the continuum data of the upper and lower sidebands were combined.

\subsection{Single-dish data}
The 1.3 mm continuum data of the $\rho$-Oph A region were taken with the IRAM 30m telescope by \citet{Motte1998}. The observational details are described in \citet{Motte1998}. The continuum image has an effective beam resolution of 13\arcsec and 1 sigma noise level of $\sim$10 mJy beam$^{-1}$.
The single-dish data of the C$^{18}$O (2--1) line were obtained through the JCMT science archive. The C$^{18}$O observations were conducted on May 3, 1999  using the RxA3 receiver. The backend was a Digital Autocorrelating Spectrometer that was configured to have a spectral resolution of 82.1 kHz (0.1067 km s$^{-1}$). The FWHM beam size of the JCMT at a frequency of the C$^{18}$O line is 22\arcsec. The 6\arcmin $\times$ 6\arcmin area including the entire region of the $\rho$-Oph A ridge was observed in the on-the-fly mode. The raw data were calibrated and gridded to the data cube using the ^^ ^^ ORAC-DR" pipeline \citep{Cavanagh2008} in the Starlink software. The resultant rms noise level of the data cube is $\sim$0.7 K per channel.

\subsection{Combining Single-dish and Interferometer data}
In order to fill the short spacing information that was not sampled by the interferometer, the SMA continuum data were combined with the IRAM 30m data, and the SMA C$^{18}$O data cube were combined with the JCMT data cube.  We used the MIRIAD package and followed the procedure described in \citet{Takakuwa2003}, which is based on the description of combining single-dish and interferometric data in \citet{Vogel1984}. The combined maps were deconvolved using the CLEAN-based algorithm, MOSSDI \citep{Sault1996}. The synthesized beam is 5\farcs2$\times$3\farcs1 (P.A. = 6.1$\arcdeg$) for the continuum map and that is 4\farcs9$\times$3\farcs4 (P.A. = 6.1$\arcdeg$) for the C$^{18}$O map.

\subsection{The SMA observations toward SM1}
The additional SMA observations toward one of the submillimeter sources, SM1, were done on Mar. 7 and 9 in 2016 in the compact configuration. The two tracks were conducted with different frequency settings, one including the CH$_3$OH (5$_k$--4$_k$) lines at 241.8 GHz, and the other including the H$_2$CO (3$_{1,2}$--2$_{1,1}$) line at 225.698 GHz, H$_2$CO
(3$_{1,3}$--2$_{1,2}$) line at 211.211 GHz and C$^{17}$O (2--1) line at 224.714 GHz.  The 4 GHz bandwidth in each sideband was covered by 48 chunks of 104 MHz bandwidth. The spectral resolution of the chunks with the CH$_3$OH, H$_2$CO, and C$^{17}$O lines was 204 kHz (0.264 km s$^{-1}$) per channel. The bright quasar, 3c273, was observed as a bandpass calibrator, and the nearby quasar, 1517-243, was used as phase and amplitude calibrators. The flux calibrators of the first track were Callisto and Ganymede, and those of the second track were Europa and MWC349A. The visibility data were calibrated using the MIR package, and imaged using the MIRIAD package. 
%-- beam size and rms noise level of the line data
The synthesized beam sizes are 3\farcs3$\times$2\farcs4 for the CH$_3$OH data cube, 4\farcs0$\times$2\farcs7 for the H$_2$CO (3$_{1,3}$--2$_{1,2}$) data cube, 3\farcs3$\times$2\farcs4 for the H$_2$CO (3$_{1,2}$--2$_{1,1}$) data cube and 3\farcs7$\times$2\farcs5 for C$^{17}$O data cube. The 1$\sigma$ rms noise level is $\sim$0.09 Jy beam$^{-1}$ per channel in CH$_3$OH and H$_2$CO data cube, and that of the C$^{17}$O data cube is $\sim$0.07 Jy beam$^{-1}$. Only the two CH$_3$OH lines with the lowest energy levels (5$_0$--$4_0$ A, $E_{\rm up}$= 34.8 K and 5$_{-1}$--4$_{-1}$ E, $E_{\rm up}$= 40.4 K) were detected. Both H$_2$CO lines (3$_{1,3}$--2$_{1,2}$, $E_{\rm up}$= 32.1 K and 3$_{1,2}$--2$_{1,1}$, $E_{\rm up}$= 33.4 K) were detected in the SMA observations, but we only present the H$_2$CO (3$_{1,3}$--2$_{1,2}$) line, since the lines having similar energy levels show similar structures. 
The continuum data were obtained by averaging the line-free chunks of both sidebands. The data of two sidebands were combined in order to improve the signal-to-noise ratio. In addition, the continuum visibility data of the two tracks are also combined. The synthesized beam and the 1$\sigma$ rms noise level of the continuum map are 3\farcs9$\times$2\farcs9 (P.A. = -9.1$\arcdeg$) and 2.7 mJy beam$^{-1}$, respectively.

\begin{figure*}
\epsscale{0.95}
%\plottwo{cont_smamotte.jpg}{c18o_mom0.jpg}
\plotone{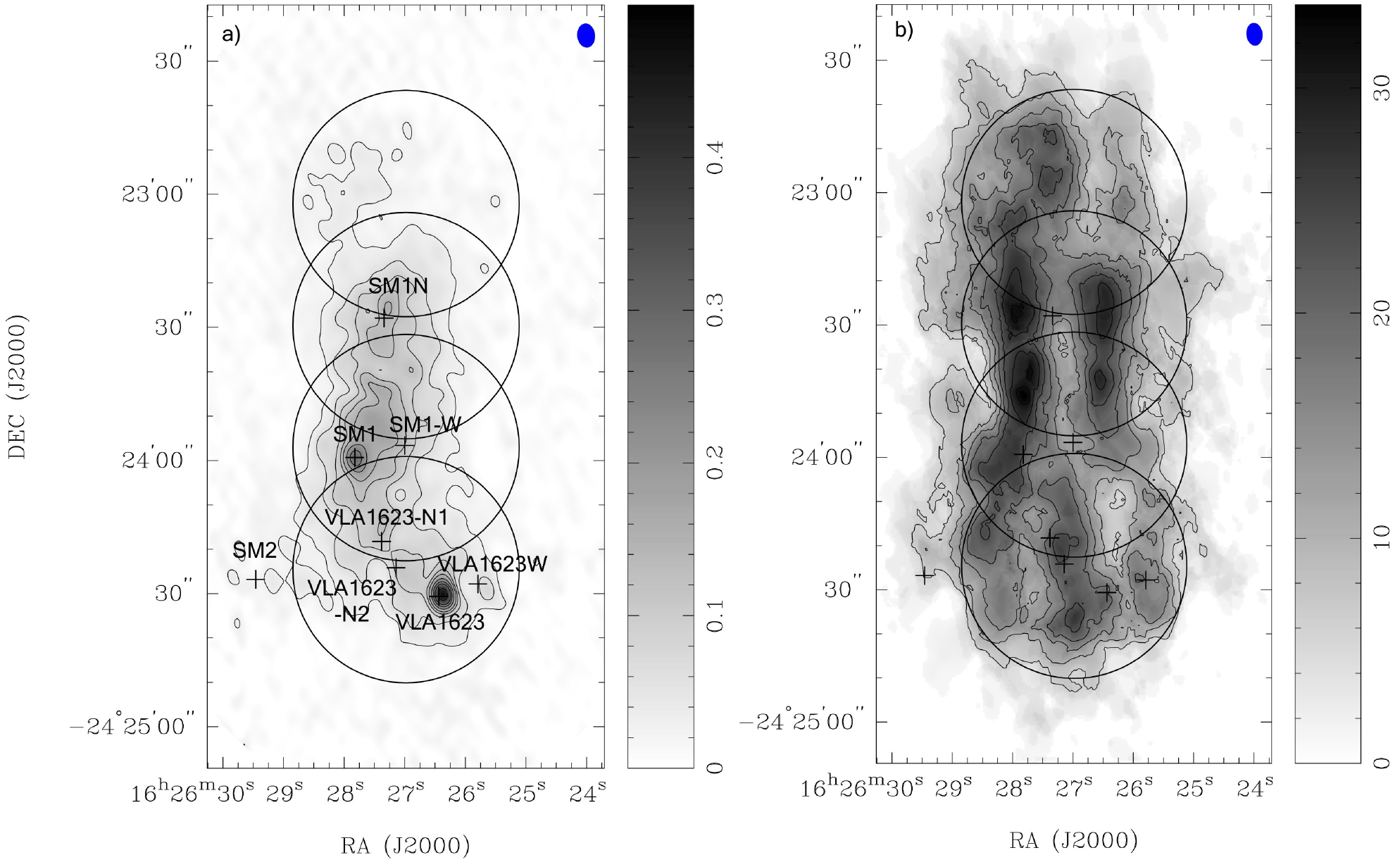}
\caption{(a) 1.3 mm Continuum map obtained by combining the IRAM 30m data and the SMA data of 4-pointing mosaic. The 1$\sigma$ rms noise level of the map is 8.5 mJy beam$^{-1}$. Solid contours denote 30, 60, 90, 120, 150, 200, 250, 300 and 400 mJy beam$^{-1}$. Gray scale starts from 0 to 470 mJy beam$^{-1}$. The cross signs except SM2 and SM1N are the 1.3 mm peak determined from the maps presented in Figure \ref{fig:cont_sm1}. The locations of SM2 and SM1N are from \citet{Motte1998}. The ellipse at top right corner denotes the beam size of 5\farcs2$\times$3\farcs7. Solid circles denote the field of views of the SMA in four pointing mosaic observations. (b) C$^{18}$O (2--1) integrated intensity map derived from JCMT 15m telescope and the SMA data. The 1$\sigma$ rms noise level of the map is 2.1 Jy beam$^{-1}$ km s$^{-1}$. Solid contours denote 3, 5, 7, 9, 11, 13 and 15 $\times$ 1$\sigma$ rms level. The range of the gray scale is 0--33.7 Jy beam$^{-1}$ km s$^{-1}$. 
%The value here are the sum of CLEANed data which intensity exceeding 2$\sigma$ rms per channel. 
The ellipse at top right corner denotes the beam size of 4\farcs9$\times$3\farcs4. }
\label{fig:cont_4point}
\end{figure*}

\begin{figure}
\epsscale{1.0}
\plotone{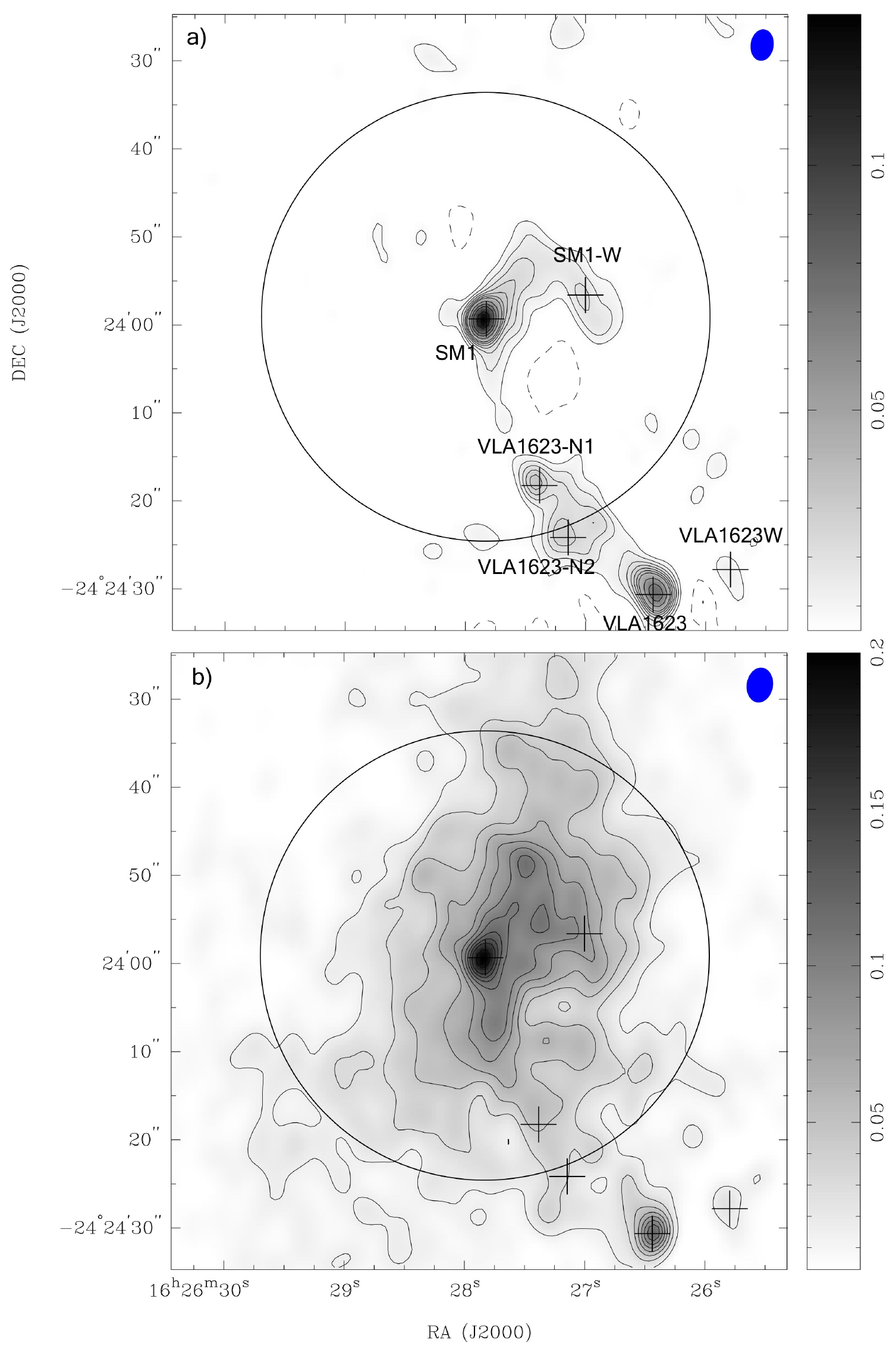}
%\plottwo{cont_sm1_labeled.jpg}{cont_sm1_combined.jpg}
\caption{(a) The SMA map of the 1.3 mm continuum centered at SM1. The 1$\sigma$ rms noise level is 2.7 mJy beam$^{-1}$. The solid contours denote 3, 5, 7, 9, 11, 13, 15, 20, 25, 30, 40 and 50 $\times$ 1$\sigma$ level, while the dashed contour denotes -3$\sigma$ levels. The gray scale ranges from 5 to 131 mJy beam$^{-1}$. The ellipse at the top right denotes the synthesized beam of 3\farcs5$\times$2\farcs5. Solid circle denotes the field of view of the SMA.
(b) The SMA + IRAM 30m map of the 1.3 mm continuum centered at SM1. The 1$\sigma$ rms noise level of the map is 3 mJy beam$^{-1}$. Solid contour denote 5, 10, 15, 20, 25, 30, 35, 40, 45, 50, 60, 70, 80 and 90 $\times$ 1$\sigma$ level. The gray scale ranges from 5 to 200 mJy beam$^{-1}$. The ellipse at top right corner denotes the synthesized beam of 3\farcs9$\times$2\farcs9 . The cross signs denote the continuum peaks derived from the Gaussian fitting. }
\label{fig:cont_sm1}
\end{figure}

\section{Results and analysis}

\subsection{1.3 mm continuum}
Figure \ref{fig:cont_4point}a shows the 1.3 mm continuum emission from $\rho$-Oph A region obtained by the combination of the IRAM 30 m data and the SMA interferometer data. The continuum emission shows the N-S ridge similar to the single-dish map presented in \citet{Mezger1992}, \citet{Andre1993} and \citet{Motte1998}. There are three major components, VLA1623, SM1, and SM1N. Our high resolution map having an angular resolution of $\sim$5\arcsec reveals that each source has internal structure. The brightest source in this region is VLA1623, a well-studied class 0 protostar \citep[e.g.][]{Andre1993}. The recent studies by \citet{Murillo2013a} and \citet{Chen2013} showed that VLA1623 consists of three sources, VLA1623A, B and W, which are suggested to be a triple non-coeval system. In our continuum map, VLA1623A and B, having a separation of $\sim$1\arcsec, are not spatially resolved. On the other hand, VLA1623W is identified as a separate peak. 

The emission from the second brightest source, SM1, shows a well-defined peak. Recent ALMA observations at 359 GHz \citep{Friesen2014} have revealed that this source is very compact, having a size of only 0.31\arcsec ( $\sim$42 au at 137 pc). The compact source is surrounded by the extended structure having an elongation toward the north-western direction.
In the SM1 region, \citet{Nakamura2012} reported three condensations. 
The brightest one, a1, corresponds to the continuum peak of our map. The other two, a2 and a3, are not clearly seen in our map, probably because of the low intensity contrast of these components and the lower angular resolution of our map. 
%Besides, there are two sources between SM1 and VLA1623, labeled as VLA1623-N1 and N2. 
VLA1623 and SM1 are connected by the curved emission ridge, having two local peaks labeled as VLA1623-N1 and N2.
In contrast to SM1 and VLA1623, the northern source SM1N is spatially extended and contains no compact component, which is consistent with the 359 GHz results of \citet{Friesen2014}.
The emission from SM2 is less significant, because this source is located beyond the area covered by our 4-pointing mosaic. The 3mm continuum map in \citet{Kirk2017} have the similar structure as our continuum map; both the extended filament around north-east of SM1 and VLA1623-N1 are clearly seen in their map.

%The SMA continuum map at SM1 is presented in Figure 2a. 
Figure \ref{fig:cont_sm1}a shows the continuum map of the SM1 region observed with the SMA in 2016.
With higher sensitivity thanks to the wider bandwidth, two sources, VLA1623-N1 and N2, are clearly seen in this map.
The SMA data are also combined with the IRAM 30m data (Figure \ref{fig:cont_sm1}b)
%We also derive the combined continuum map at same region in Figure 2b by combining the IRAM 30 m data and the new SMA interferometer data. 
%The sources such as VLA1623, SM1, VLA1623-N1 and N2 can be identified in newest SMA continuum map. but VLA1623W is not obvious due to its low intensity. SM1 is the brightest source in the field and there are also elongated structure toward north-western direction. 
The combined map recovers well the extended emission around the SM1 peak.
The continuum emission extends to the north of SM1, which is consistent with the 340 GHz results of \citet{Nakamura2012}.
However, the small scale condensations a2 and a3 reported by \citet{Nakamura2012} are not clearly seen in our maps.
Instead, our SMA map shows the local intensity maximum to the west of SM1, which is labeled as SM1-W.
%The local intensity maxima to the north and northwest of the SM1 peak\edit1{, labeled as SM1-W,} imply the presence of small-scale condensations suggested by \citet{Nakamura2012}.
%However, the locations of these peaks do not coincide with those of a2 and a3 in \citet{Nakamura2012}.
Instead of being obvious in the SMA map (Figure \ref{fig:cont_sm1}a), VLA1623-N1, -N2 and SM1-W are not significant in the combined map (Figure \ref{fig:cont_sm1}b) because of the rather high-level emission from the extended component.

\begin{deluxetable*}{c c c c c c c c c }
\tablecaption{The fitting parameters of each submillimeter source}
\tablehead{\colhead{Source} & \colhead{R.A.} & \colhead{Dec.} & \colhead{S$_p$} & \colhead{S$_t$ } & \colhead{FWHM} & \colhead{P.A.} & \colhead{Mass\tablenotemark{a}} & \colhead{M$_{BE}$} \\ 
   \colhead{}   & \colhead{(J2000)} & \colhead{(J2000)}&  \colhead{(Jy beam$^{-1}$)} & \colhead{(Jy)}   &  \colhead{(arcsec)} & \colhead{(degrees)}& (M$_{\odot}$) & \colhead{(M$_{\odot}$)}}
\startdata
VLA1623 & 16:26:26.44 & -24:24:30.65 & 0.50 & 0.86 & 3\farcs$2\times$2\farcs4 & 7.3 & 0.41 & 0.010 \\
VLA1623-N1\tablenotemark{*} &  16:26:27.38& -24:24:18.26 & 0.049 & 0.11 & 4\farcs7$\times$1\farcs8 & 44.3  & 0.05 & 0.014\\
VLA1623-N2\tablenotemark{*} & 16:26:27.15 & -24:24:24.17 & 0.06 & 0.31 & 6\farcs8$\times$5\farcs3 & 60.1 & 0.15 & 0.02\\
VLA1623W & 16:26:25.80 & -24:24:27.81 & 0.14 & 0.32 & 5\farcs5$\times$2\farcs4 & 14.4 & 0.15 & 0.017\\
SM1 & 16:26:27.83 & -24:23:59.34 & 0.19 & 0.59 & 5\farcs8$\times$4\farcs2 & -9.3 & 0.28 & 0.018 \\
SM1-W\tablenotemark{*} & 16:26:27.00 & -24:23:56.61 & 0.02 & 0.17 & 12\farcs2$\times$4\farcs0 & 36.4 & 0.08 & 0.038\\
\enddata
\tablenotetext{*}{The parameters for VLA1623-N1, VLA1623-N2 and SM1-W are determined from the SMA map (Figure \ref{fig:cont_sm1}a) only. }
\tablenotetext{a}{
%The 1$\sigma$ uncertainty of mass estimation is $\sim$ 10$\%$ of mass estimated from the 1$\sigma$ error of peak emission.
The uncertainty in the mass calculation derived from the 1$\sigma$ error in the measured flux is $\sim$ 10$\%$.}
\label{table:cont}
\end{deluxetable*}

%which means the ridge could harbors smaller scale fragments but they are not as bright as the extended component surrounding SM1.

The coordinates, peak flux density, total flux, size, and position angle of each source were determined from the single Gaussian component fitting using the MIRIAD task IMFIT.
The size and position angle of each source are derived after deconvolved with the beam.
The parameters of VLA1623-N1, -N2 and SM1-W are derived from the SMA map shown in Figure \ref{fig:cont_sm1}a, while those of VLA1623, VLA1623W, and SM1 are from the combined map shown in Figure \ref{fig:cont_sm1}b.
The derived parameters are listed in Table 1.
%The sources including VLA1623, VLA1623W, SM1, VLA1623-N1 and N2 are fitted by MIRIAD task IMFIT with Gaussian model. The physical property of VLA1623-N1 and N2 are derived from the SMA map centered at SM1 (Fig 2a), and those of VLA1623, VLA1623W and SM1 are derived from the combined map (Fig 2b). 
%Their parameters such as coordinate and peak intensity are listed in Table 1.

\subsection{C$^{18}$O (2--1)}
The integrated intensity map of the C$^{18}$O obtained by combining the JCMT data and the SMA data is presented in Figure \ref{fig:cont_4point}b. 
Although the C$^{18}$O emission  comes from the N-S ridge of the  $\rho$-Oph A, its spatial distribution is significantly different from that of the continuum; the C$^{18}$O emission is bright in the eastern and western edges of the ridge, and rather faint at the center of the ridge where the continuum emission is bright. 

In the northern part, the anti-correlation between the C$^{18}$O and continuum is significant; the continuum peak SM1N is located between two bright ridges of the C$^{18}$O.
In the southern part, there is a third ridge of C$^{18}$O emission connecting SM1 and VLA1623.
However, the C$^{18}$O does not follow the curved ridge traced by the 1.3 mm continuum. In addition, locations of the C$^{18}$O peaks do not coincide with the continuum peaks.
On the other hand, the C$^{18}$O peak coincides with the continuum peak at VLA1623W.

\begin{figure}
%\plotone{c18on2h.jpg}
\plotone{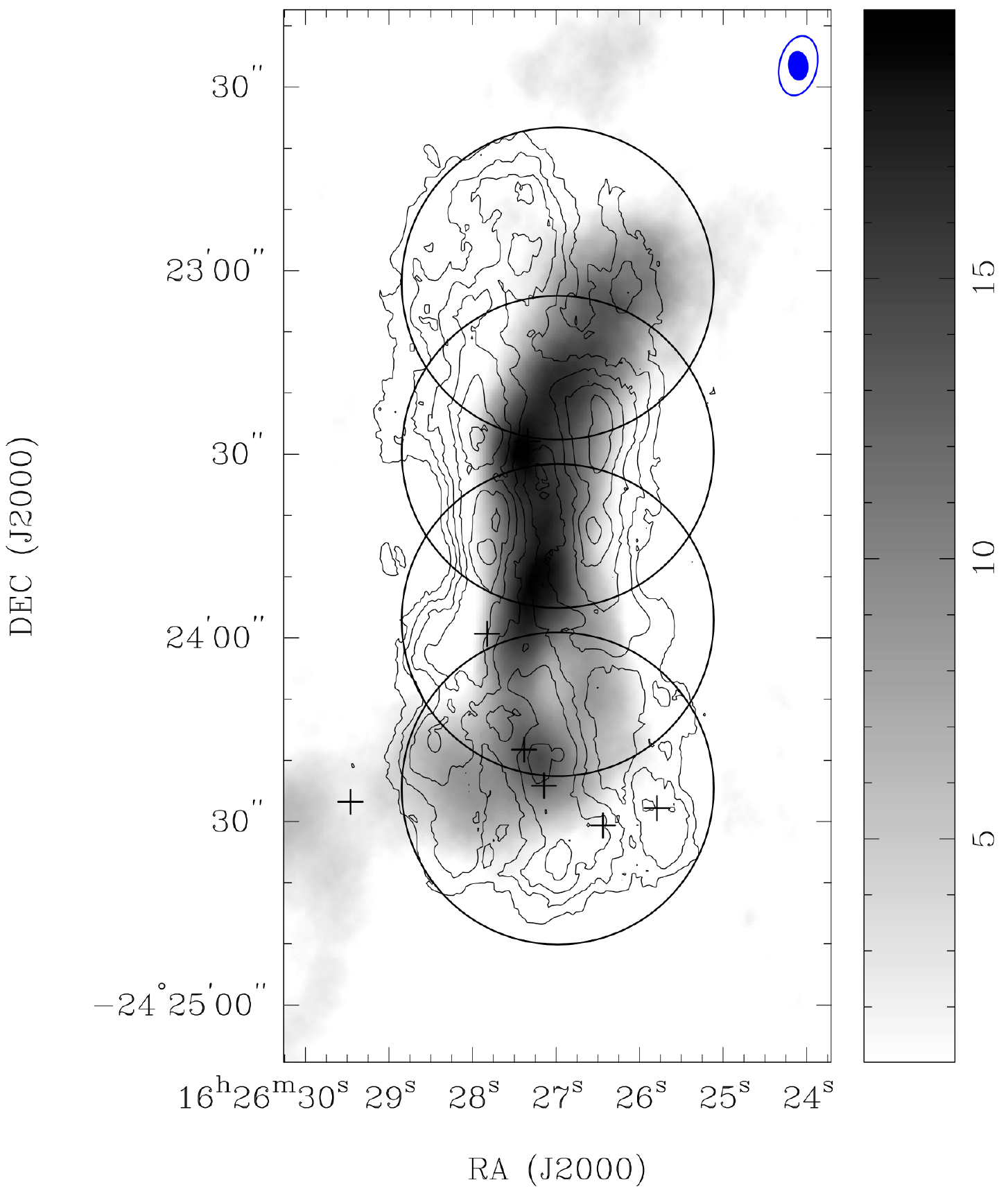}
\caption{A comparison between C$^{18}$O (contours) and N$_2$H$^+$ (gray scale) integrated intensity maps. The N$_2$H$^+$ map is obtained from \citet{DiFrancesco2004}. The rms noise level, contours and resolution of the C$^{18}$O map are the same as in Figure \ref{fig:cont_4point}b. For the N$_2$H$^+$ map, the 1$\sigma$ rms noise level is 1.0 Jy beam$^{-1}$ km s$^{-1}$ and the gray scale ranges from 1.0 to 19.8 Jy beam$^{-1}$ km s$^{-1}$. The hollow ellipse at the top right corner denotes the beam size of the N$_2$H$^+$ map, 9\farcs9$\times$6\farcs2.}
\label{fig:c18on2h}
\end{figure}

Since our C$^{18}$O map contains the short-spacing information obtained with the JCMT, the lack of  bright emission at the center of the ridge is not the effect of spatial filtering of the interferometer.
The CO and $^{13}$CO lines observed in the $\rho$-Oph molecular cloud often show the self-reversal line profile, which could produce intensity drop at the line of sight at high optical depth.
However, the observed C$^{18}$O line does not show such a self-reversal profile.
Such an anti-correlation between the C$^{18}$O and continuum was not clear in the previous single-dish observations \citep{Liseau2010,White2015} due to their angular resolutions. 
Recently, \citet{Liseau2015} deconvolved their C$^{18}$O (3--2) image, and improved the angular resolution from 19$\arcsec$ to 7$\arcsec$.
Their deconvolved C$^{18}$O image also reveals the anti-correlation with the continuum distribution.
Moreover, by comparing our C$^{18}$O integrated intensity map with the N$_2$H$^+$ map from \citet{DiFrancesco2004}, we found that the most of the N$_2$H$^+$ emission comes from the region between two C$^{18}$O ridges (Figure \ref{fig:c18on2h}). The highest N$_2$H$^+$ emission occurs across SM1N and SM1, where the C$^{18}$O emission is missing. The spatial distribution of the C$^{18}$O is anti-correlated with that of N$_2$H$^+$ except the southern ridge including VLA1623-N1 and VLA1623-N2.

\begin{deluxetable*}{c c c c c c c c }
\tablecaption{The column density and abundance of each submillimeter source}
\tablehead{\colhead{Source} &\colhead{$N_{H_2}$} & \colhead{$N_{C^{18}O}$} & \colhead{$N_{N_2H^+}$} & \colhead{$X_{C^{18}O}$} & \colhead{$X_{N_2H^+}$} & \colhead{$X_{C^{17}O-SMA}$} & \colhead{$X_{C^{18}O-SMA}$}\\ 
   \colhead{}   & \colhead{($10^{23}$cm$^{-2}$)} & \colhead{($10^{16}$cm$^{-2}$)}   &  \colhead{($10^{12}$cm$^{-2}$)} & \colhead{($10^{-8}$)}& \colhead{($10^{-11}$)} & \colhead{(10$^{-9}$)} & \colhead{(10$^{-8}$)}}
\startdata
VLA1623 & 5.8 & 2.5  & 7.8 & 4.3& 1.4 & 3.6\tablenotemark{b} & 1.3\tablenotemark{c}\\
VLA1623-N1 & 2.7 & 5.2 & 54.2 & 24.4 & 20.2 & 2.4\tablenotemark{b} & 0.88\tablenotemark{c} \\
VLA1623-N2 & 2.4 & 4.5 & 60.1 & 18.8 & 25.1 & 6.0\tablenotemark{b} & 2.2\tablenotemark{c}\\
VLA1623W & 1.5 & 2.1 & $<$ 6\tablenotemark{a} & 14.2& $<$3.9\tablenotemark{a} & $<$1.7\tablenotemark{a}\tablenotemark{b} & $<$0.62\tablenotemark{c}\\
SM1 & 5.4 & 10.2 & 17.0 & 19.1& 3.2  & 2.4\tablenotemark{b} & 0.88\tablenotemark{c}\\
SM1-W & 2.2 & 6.6 & 65.5 & 3.8 & 20.5 & $<$5.2\tablenotemark{a}\tablenotemark{b} & $<$1.9\tablenotemark{c} \\
\enddata
\tablenotetext{a}{The flux of the N$_2$H$^+$(C$^{17}$O) emission is below the 3$\sigma$ noise level.}
\tablenotetext{b}{The C$^{17}$O abundance is calculated from the SMA data only.}
\tablenotetext{c}{The C$^{18}$O abundance is derived by multiplying the C$^{17}$O abundance estimated from the SMA data by the $^{18}$O/$^{17}$O abundance ratio, 3.65 \citep{Penzias1981}.}

\end{deluxetable*}

\begin{figure}
\epsscale{1}
%\plotone{c18o_peak_tb.jpg}
\plotone{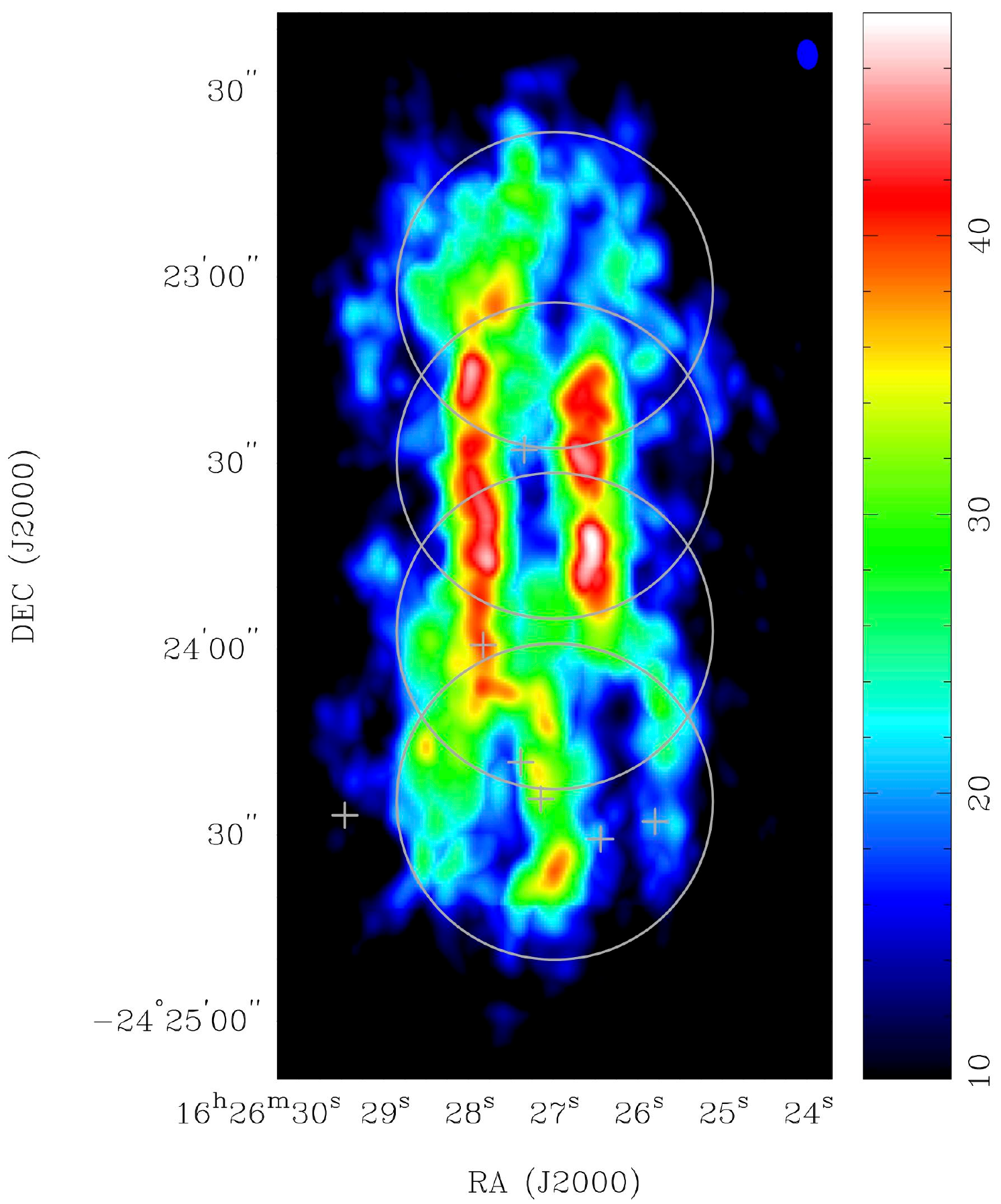}
\caption{The peak intensity map of the C$^{18}$O in the $\rho$-Oph A ridge. The color scale ranges from 10 K to 48 K. The resolution and cross symbols are the same as those in Figure \ref{fig:cont_4point}b.}
\label{fig:temp}
\end{figure}

The C$^{18}$O peak intensity map shown in figure \ref{fig:temp} indicates that the prominent C$^{18}$O ridges have high brightness temperature of $>$40 K; whereas, in the rest region, the temperature is roughly 25 K or below. 
A notable thing is that, except SM1, where the temperature is roughly 40 K, major dust condensations are located in the regions with lower brightness temperature; VLA1623-N1 and VLA1623-N2 are in the region with $\sim$30 K, and SM1N, VLA1623, and VLA1623W are in the regions of $\sim$25K.

\begin{figure}
\epsscale{1.1}
%\plotone{c18o_cm.jpg}
\plotone{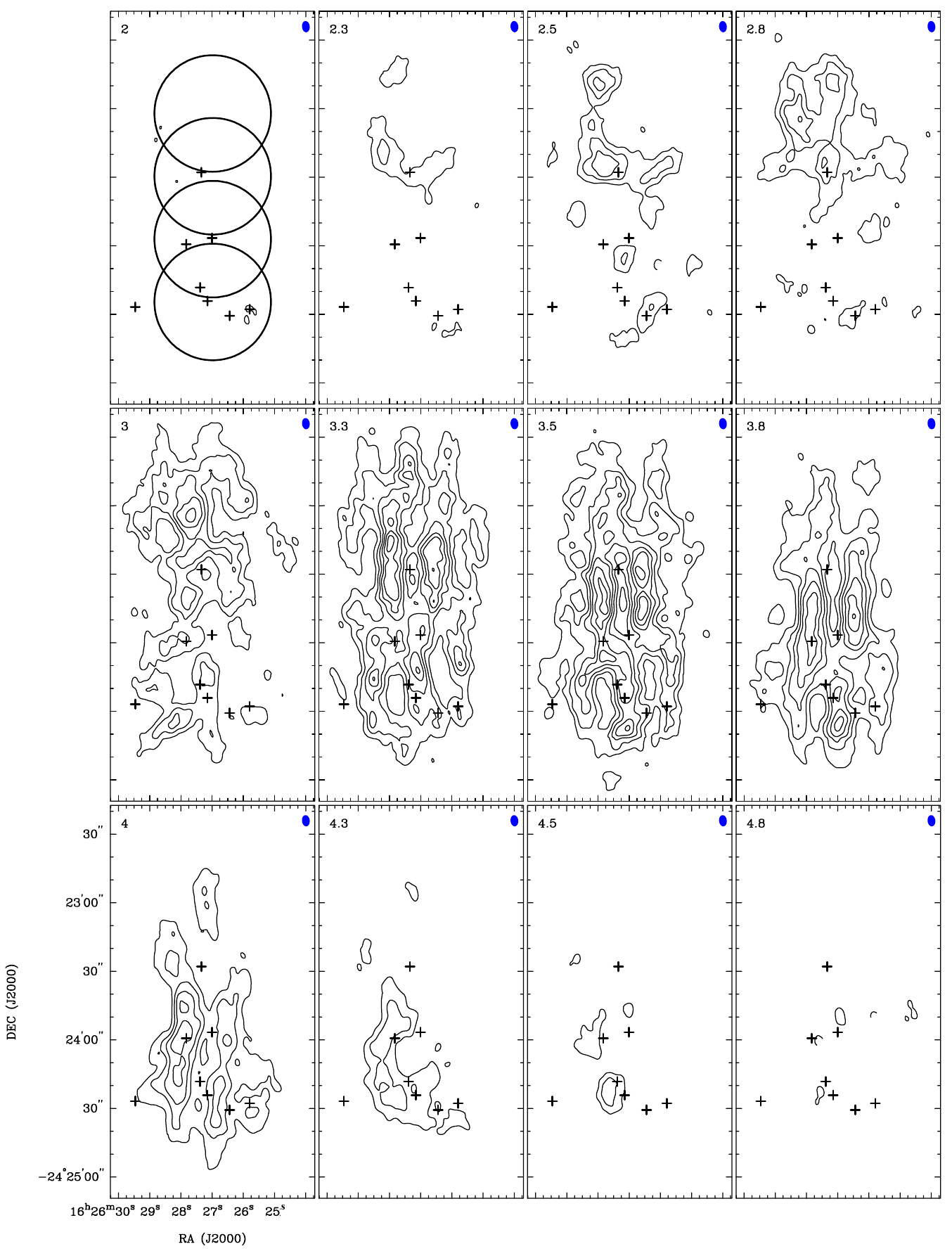}
\caption{Velocity-channel maps of the C$^{18}$O. 1$\sigma$ rms noise level of the map is 2.1 Jy beam$^{-1}$. The solid contours denote 3, 5, 7, 9, 11 and 13 $\times$ 1$\sigma$ level, while the dashed contour denotes -3$\sigma$ level. The resolution and cross symbols are the same as those in Figure \ref{fig:cont_4point}b.}
\label{fig:channel}
\end{figure}

\begin{figure}
%\plotone{velplot.jpg}
\epsscale{0.9}
\plotone{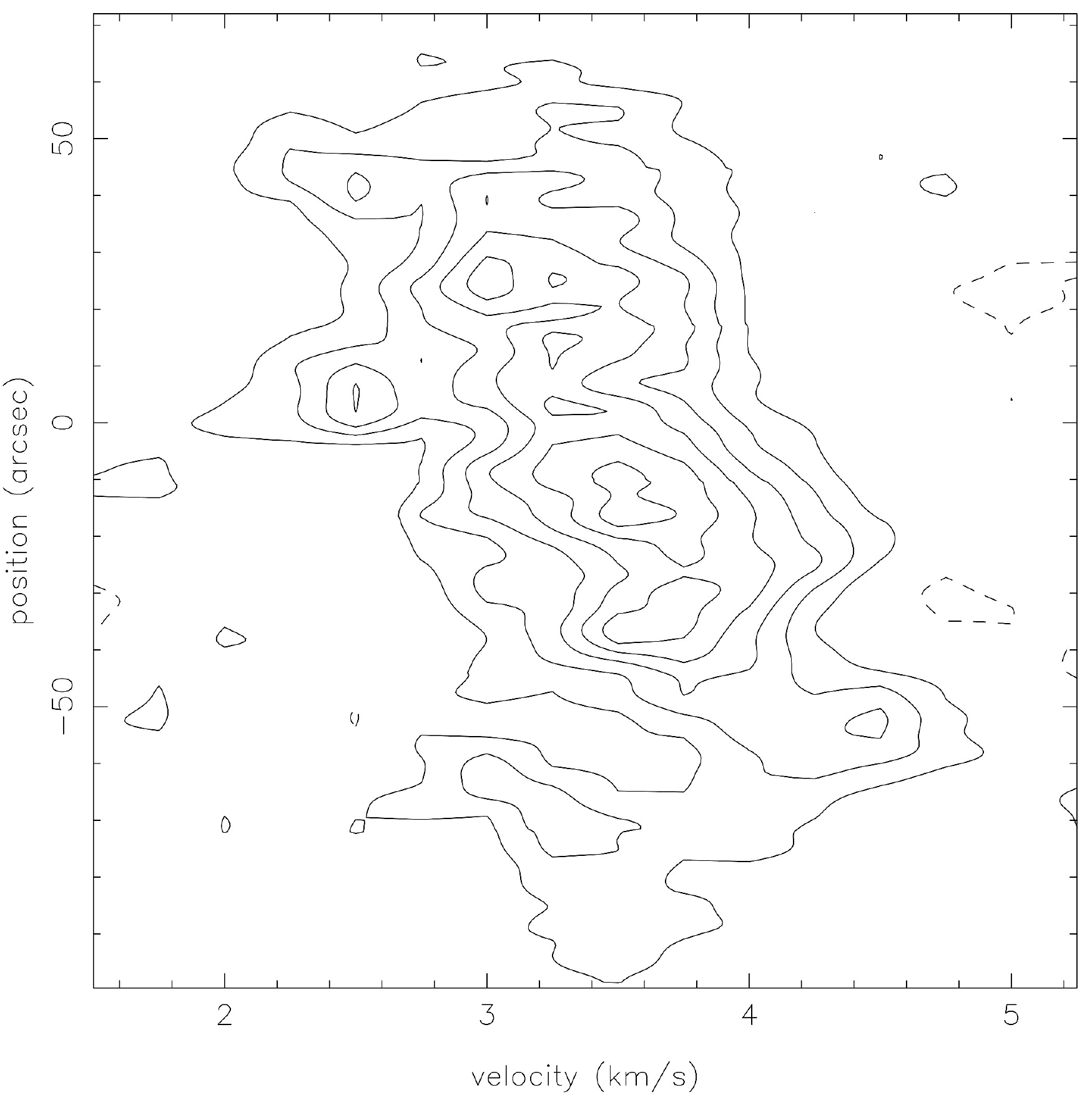}
\caption{Position-Velocity diagram of the C$^{18}$O along the N-S direction centered at R.A= 16:26:26.98 and Dec=-24:23:29.60. The solid contours denote 4.32, 8.64, 13.0, 17.3, 21.6, 25.9 Jy beam$^{-1}$.}
\label{fig:velplot}
\end{figure}

The channel map (Figure \ref{fig:channel}) and the position-velocity map along the N-S cut through R.A. = 16$^h$26$^m$27$^s$ (Figure \ref{fig:velplot}) show that there is a significant velocity gradient along the $\rho$-Oph A ridge.
%\textcolor{blue}{The velocity structure of the $\rho$-Oph A ridge \textcolor{red}{is shown} in the channel maps (Figure \ref{fig:channel}) and the position-velocity map through R.A. = 16:26:27 (Figure \ref{fig:velplot})\textcolor{red}{. \sout{is consistent with those of C$^{18}$O(3-2) presented in}}. 
The velocity centroid in the northern part of the ridge is at V$_{\rm LSR}$ $\sim$ 3.1 km s$^{-1}$, and gradually changes to $\sim$ 3.7 km s$^{-1}$ at the position of SM1, which is consistent with the C$^{18}$O (3--2) results of \citet{Liseau2010}.
The southern redshifted component shows an arc-like feature at $\sim$4.3 km s$^{-1}$. This arc-like feature follows the eastern edge of the continuum arc connecting SM1 and VLA1623 (see Panel 10 of Figure \ref{fig:channel}).
\citet{Bergman2011} have also shown that the chemical properties are different between the northern part and southern part of the ridge.
The deuterated species like HDCO, D$_2$CO and N$_2$D$^+$ are abundant in the southern part with $V_{\rm LSR}{\sim}$3.7 km s$^{-1}$, but are deficient in the northern part with $V_{\rm LSR}{\sim}$3.1 km s$^{-1}$.

\begin{figure}
\plotone{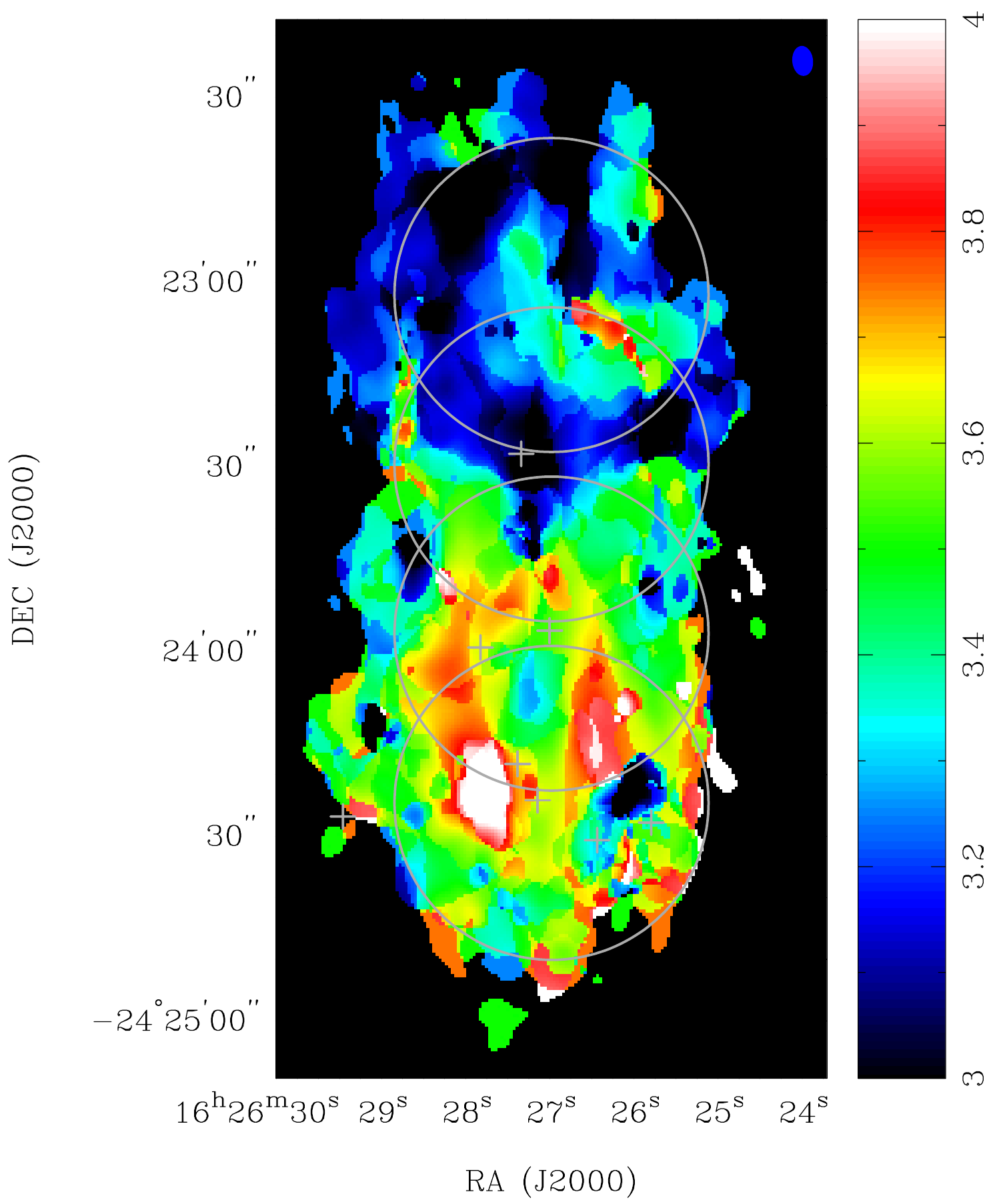}
\caption{The velocity centroid (moment 1) map of the C$^{18}$O in the $\rho$-Oph A ridge. The color scale ranges from $V_{\rm LSR}$= 3 km s$^{-1}$ to 4 km s$^{-1}$. The resolution and cross symbols are the same as those in Figure \ref{fig:cont_4point}b.}
\label{fig:mom1}
\end{figure}

The abrupt velocity change near SM1N in moment 1 map (Figure \ref{fig:mom1}) is likely to be originated from the third component at V$_{\rm LSR}$ $\sim$2.9 km s$^{-1}$, which is clearly seen in the H$_2$CO position-velocity and profile maps in \citet{Bergman2011}. This component is not clearly seen in the position-velocity map of C$^{18}$O (3--2) in \citet{Bergman2011}, while clearly seen in our C$^{18}$O (2--1) position-velocity map. Furthermore, the 2.9 km s$^{-1}$ component has the C-shaped pattern seen in the mom1 map and channel map, and might be related to the methanol rich gas, which also shows C-shaped morphology in larger scale \citep{Bergman2011,Garay2002}. Overall, it is likely that the $\rho$-Oph A ridge consists of three components with different velocities and chemical properties; the northern component at 3.1 km s$^{-1}$ which is deficient in deuterated species, the southern deuterium-rich component at 3.7 km s$^{-1}$, and the north-western component at 2.9 km s$^{-1}$ having high methanol abundance.

\subsection{C$^{18}$O and N$_2$H$^+$ Abundance}

\begin{figure}
%\plotone{c18o_abundance2.png}
\plotone{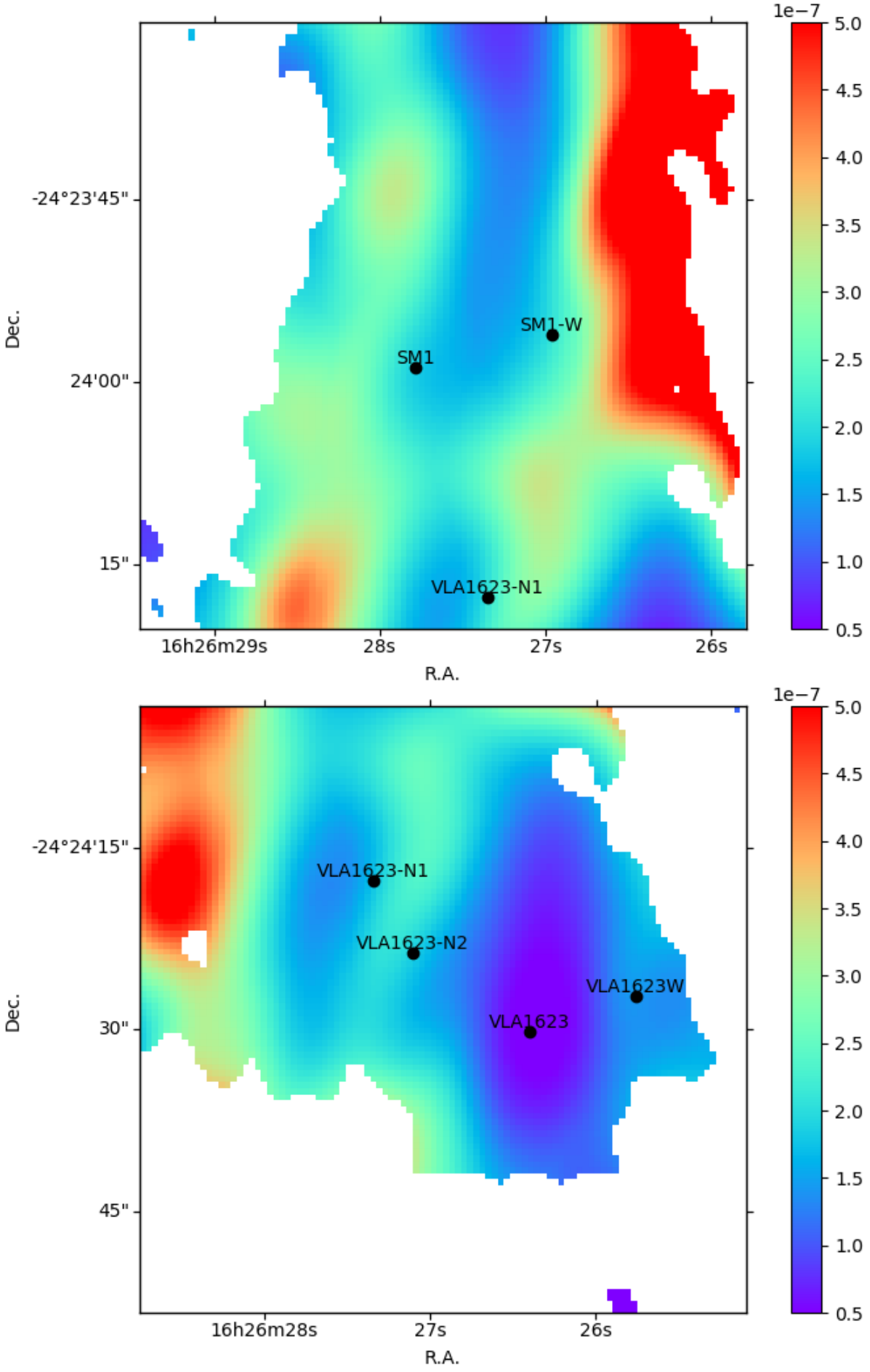}
\caption{Fractional abundance of the C$^{18}$O derived from dividing N(C$^{18}$O) by N(H$_{2}$). The color scale ranges from 2$\times$10$^{-8}$ to 5$\times$10$^{-7}$. The area where continuum emission is below 3$\sigma$ rms noise levels are clipped. The map is restricted to two 50\arcsec$\times$50\arcsec small regions where optical depths of the C$^{18}$O could be calculated.}
\label{fig:c18o_abundance}
\end{figure} 

\begin{figure}
%\plotone{n2h_abundance2.png}
%\plotone{n2h_abundance1.png}
\plotone{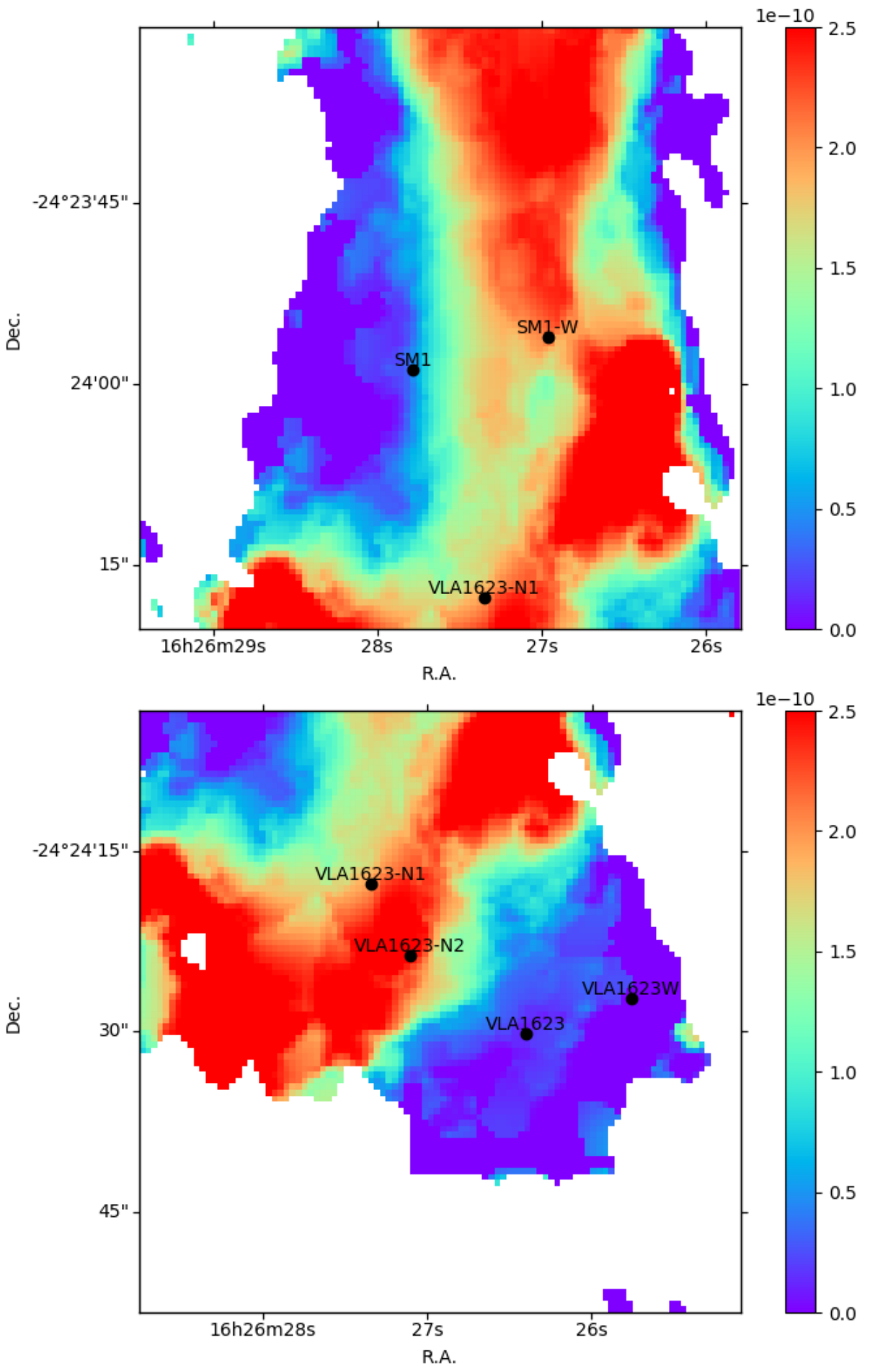}
\caption{Fractional abundance of the N$_2$H$^+$ derived from dividing N(N$_2$H$^+$) by N(H$_{2}$). The color scale ranges from 0 to 1$\times$10$^{-10}$. The area where continuum emission is below 3$\sigma$ rms noise levels are clipped. 
%The map is restricted to two 50\arcsec$\times$50\arcsec small regions where optical depths could be calculated.
The areas in the two panels are the same as those of Figure \ref{fig:c18o_abundance}.}
\label{fig:abundance}
\end{figure} 

The column density of H$_{2}$ in $\rho$-Oph A can be determined from 1.3 mm continuum emission by the formula
\begin{equation}
N(\text{H}_2)=\frac{S_{\nu}}{\Omega_m\mu m_{\text{H}}\kappa_{\nu}B_{\nu}(T_{\text{dust}})}
\end{equation}
where S$_{\nu}$ is the 1.3 mm flux density, $\Omega_m$ is the solid angle of the beam, $\mu$ is the mean molecular weight, m$_{\rm H}$ is the mass of the atomic hydrogen, $\kappa_{\nu}$ is the dust opacity per unit mass. (assumed to be 0.005 cm$^{2}$ g$^{-1}$, which is the same value used by \citet{Motte1998} for pre-stellar clumps and cores), and B$_{\nu}(T_{\text{dust}})$ is the Plank function at the dust temperature, which is assumed to be 27 K \citep{Andre1993}.

On the other hand, by following \citet{Mangum2015}, the C$^{18}$O and N$_2$H$^+$ column densities can be derived from the local thermodynamic equilibrium (LTE) assumptions,

\begin{eqnarray}
N(C^{18}O)=  \frac{3.15 \times 10^{15}}{\theta_a(\arcsec)\theta_b(\arcsec)} (\frac{\tau}{1-e^{-\tau}}) \frac{\exp \left(\frac{15.8}{T_{\rm ex}}\right)}{\exp\left(\frac{10.54}{T_{\rm ex}}\right)-1} \nonumber\\
 \times  \left(\frac{T_{\rm ex}+0.88}{T_{\rm ex}-2.73}\right) \int S_{\nu}(\text{Jy}) dv(\text{kms}^{-1})  \text{ cm}^{-2}
\end{eqnarray}
and,
\begin{eqnarray}
N(N_2H^+)=  \frac{4.48 \times 10^{13}}{\frac{1}{9} \theta_a(\arcsec)\theta_b(\arcsec)}(\frac{\tau}{1-e^{-\tau}}) \frac{\exp \left(\frac{4.47}{T_{\rm ex}}\right)}{\exp\left(\frac{4.47}{T_{\rm ex}}\right)-1} \nonumber\\
 \times \left(\frac{T_{\rm ex}+0.75}{T_{\rm ex}-2.73}\right) \int S_{\nu}(\text{Jy}) dv(\text{km s}^{-1})\text{ cm}^{-2}
\end{eqnarray}

where $\theta_a$ and $\theta_b$ are the major and minor axes of the beam (FWHM in arcseconds), respectively, $T_{\rm ex}$ is the excitation temperature, $\tau$ is the optical depth, $S_{\nu}$ is flux density, and $dv$ is the velocity interval.
Since C$^{18}$O is not optically thin in the $\rho$-oph A region, the optical depth of the C$^{18}$O emission was estimated using the C$^{18}$O (2--1) and C$^{17}$O (2--1) data observed with the JCMT \citep{Gurney2008}. Unfortunately, the C$^{17}$O dataset covers only the 50\arcsec $\times$ 50\arcsec areas centered at SM1 and VLA1623. Therefore, we could not derive the optical depth of the C$^{18}$O in the SM1N region. We followed the method described in the appendix B of \citet{Ladd1998}. The $^{18}$O/$^{17}$O abundance ratio was assumed to be 3.65 \citep{Penzias1981}. Because the JCMT data were sampled in a 10\arcsec grid, the optical depth were calculated by interpolating the data points. The calculation showed that C$^{18}$O is optically thick in most of the $\rho$-Oph A region with the highest value of 5.5 at 5\arcsec W of SM1. As for N$_2$H$^+$ lines, \citet{DiFrancesco2004} showed that the total optical depth of the N$_2$H$^+$ line is larger than unity in the $\rho$-Oph A region. Therefore, the N$_2$H$^+$ column density was calculated from the ^^ ^^ isolated" 101-012 component at 93.176265 GHz with optically thin assumption and divided by its relative intensity, which is $\frac{1}{9}$.
The excitation temperature of the C$^{18}$O formula in each pixel was derived from the peak intensity map (Figure \ref{fig:temp}). The excitation temperature of N$_2$H$^+$ was assumed to be 15 K, based on the result of HFS (HyperFine Structure) fitting in \citet{DiFrancesco2004}. In order to calculate the abundances, the resolution of the H$_2$ column density map derived from the continuum data was adjusted to be the same as those of the C$^{18}$O map and N$_2$H$^+$ map, respectively. 
After dividing C$^{18}$O column density and N$_2$H$^+$ column density by H$_2$ columns density, we derived the abundance distribution over the $\rho$-Oph A region.

The C$^{18}$O abundance distribution in the $\rho$-Oph A region is presented in Figure \ref{fig:c18o_abundance} and the N$_{2}$H$^+$ abundance distribution in the same region is presented in Figure \ref{fig:abundance}. 
The column densities of H$_2$, C$^{18}$O and N$_2$H$^{+}$ and the abundances of C$^{18}$O and N$_2$H$^{+}$ at each source position are listed in Table 2 for comparison.  As shown in Figure \ref{fig:c18o_abundance}, the C$^{18}$O abundance decreases significantly at the central N-S ridge and the position of VLA1623. The C$^{18}$O abundances at those regions are below 2$\times$10$^{-7}$.
On the other hand, the C$^{18}$O abundance value often used in the interstellar medium is 1.7--2$\times$10$^{-7}$ \citep{Frerking1982,Wannier1980}.
The higher value of 4.8$\times$10$^{-7}$ is also suggested from the CO abundance of 2.7$\times$10$^{-4}$ \citep{Lacy1994,Jorgensen2005} and [$^{16}$O/$^{18}$O] = 560 (Wilson \& Rood 1994).
The C$^{18}$O abundance values measured in the N-S ridge and VLA1623 are a factor of 5--10 lower than the interstellar values.
On the other hand, the C$^{18}$O abundance values at SM1, VLA1623-N1, VLA1623-N2, and VLA1623-W are $\sim$2$\times$10$^{-7}$, which is comparable to that of the interstellar value.

The N$_2$H$^+$ abundance varies between different sources among $\rho$-Oph A. \citet{DiFrancesco2004}  suggest it is due to different evolutionary stage of each source. 
The N$_2$H$^+$ abundance is enhanced in the northern part of the ridge and the C-shaped region including VLA1623-N1 and VLA1623-N2.
It drops significantly to the eastern edge of the ridge and the southwestern part including VLA1623.
It should be noted that the N$_2$H$^+$ abundance is very low, $\sim$1.4$\times$10$^{-11}$ at the position of VLA1623, toward which the C$^{18}$O abundance is also the lowest.

%its  outflow activities observed from H$_2$O line \citep{Bjerkeli2012} can reduce the N2H+ locally via ion-molecule reactions. \citep{Bergin1998}

\begin{figure*}[!htb]
\epsscale{1.1}
\plotone{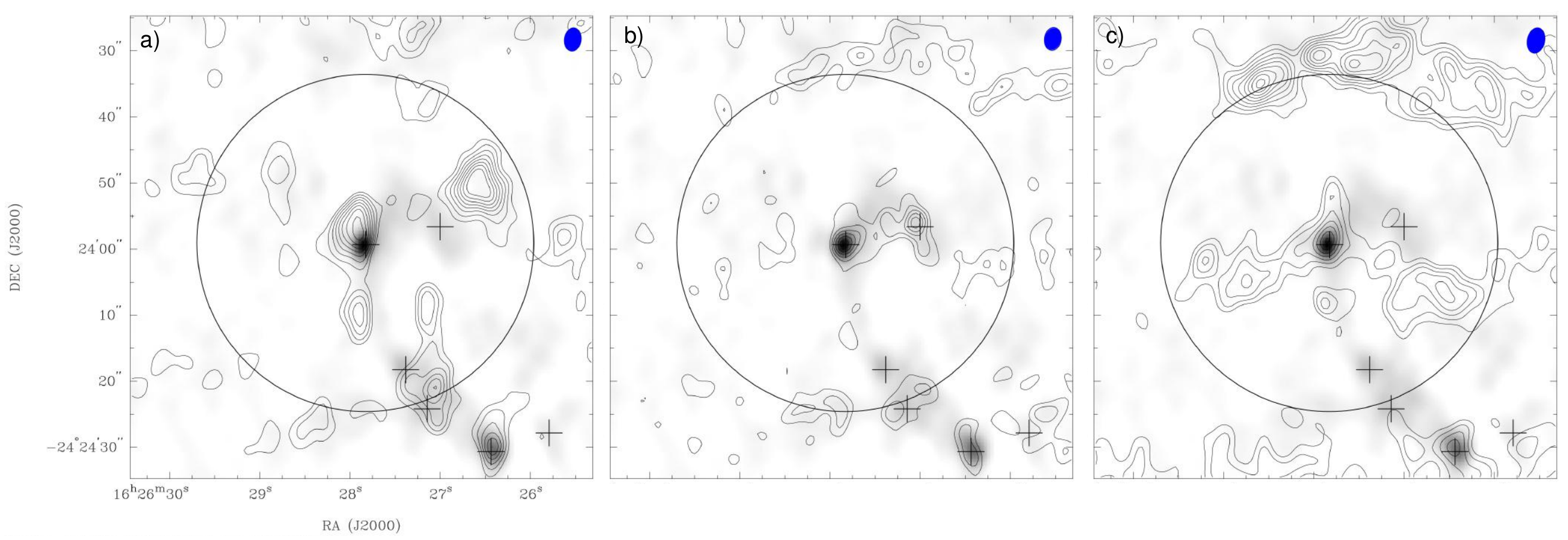}

\caption{Integrated intensity maps of the (a) C$^{17}$O (2--1), (b) CH$_3$OH ($5_0$--$4_0$A + $5_{-1}$--$4_{-1}$E), and (c) H$_2$CO (3$_{1,3}$--2$_{1,2}$) overlaid on the 1.3 mm continuum map shown in the gray scale. Contours are drawn every 2$\sigma$ levels with the lowest contours at 3$\sigma$. The 1$\sigma$ rms noise level is 0.11 Jy beam$^{-1}$km s$^{-1}$ for C$^{17}$O, 0.13 Jy beam$^{-1}$km s$^{-1}$ for CH$_3$OH, and 0.13 Jy beam$^{-1}$km s$^{-1}$ for H$_2$CO. The open circle is the field of view of the SMA. The filled ellipse at the top-right corner of each panel is the synthesized beam. The cross signs denote the positions of the continuum peaks.}
%(a)The C$^{17}$O integrated intensity contour map at SM1. The 1$\sigma$ rms noise level is 0.11 Jy beam$^{-1}$km s$^{-1}$. The solid contour denote 3, 5, 7, 9, 11, 13, 15 and 17 $\times$ 1$\sigma$ rms level. The beam size is 3\farcs62$\times$2\farcs54. The gray scale represents the continuum emission same as Figure \ref{fig:cont_sm1}a. (b) The combined CH$_3$OH ($5_0$--$4_0$A + $5_{-1}$--$4_{-1}$A) integrated intensity contour map at SM1. The 1$\sigma$ rms noise level is 0.13 Jy beam$^{-1}$km s$^{-1}$. The solid contour denote 3, 5, 7, 9, 11 and 13 $\times$ 1$\sigma$ rms level. The beam size is 3\farcs31$\times$2\farcs41. (c) The H$_2$CO (3$_{1,3}$--2$_{1,2}$) integrated intensity contour map at SM1. The 1$\sigma$ rms noise level is 0.13 Jy beam$^{-1}$km s$^{-1}$. The solid contour denote 3, 5, 7, 9, 11, 13, 15, 17 and 19 $\times$ 1$\sigma$ rms level. The beam size is 3\farcs99$\times$2\farcs72. All three maps are before the primary beam correction.}
\label{fig:sm1_line}
\end{figure*}

\begin{figure}[]
\plotone{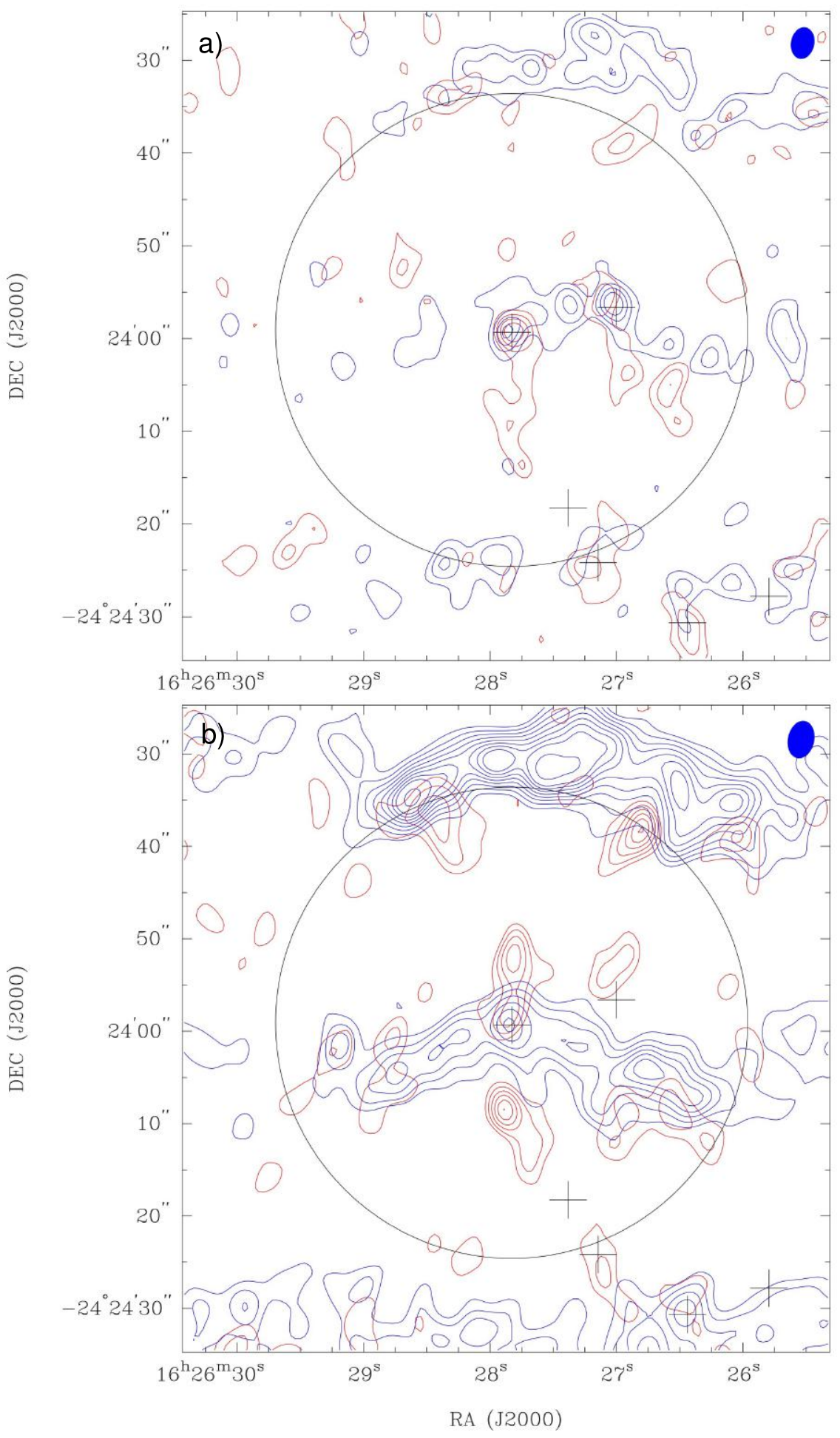}
\caption{(a) Integrated intensity maps of the blue- and redshifted components of the CH$_3$OH ($5_0$--$4_0$ A + $5_{-1}$--$4_{-1}$ E) emission. The integrated velocity ranges are 1.8--3.3 km s$^{-1}$
(blue) and 3.5--4.5 km s$^{-1}$ (red). The contours are every 2$\sigma$ with the lowest contours at 3$\sigma$. The 1$\sigma$ rms noise levels are 0.035 (red) and 0.042 (blue) Jy beam$^{-1}$km s$^{-1}$. The ellipse at top right is the beam size of 3\farcs3$\times$2\farcs4. (b) Integrated intensity maps of the blue- (2--3.5 km s$^{-1}$) and redshifted (3.8--5 km s$^{-1}$) H$_2$CO (3$_{1,3}$--2$_{1,2}$) emission.
Contour levels are from 3$\sigma$ in steps of 2$\sigma$. The 1$\sigma$ rms noise levels are 0.053 (red) and 0.058 (blue) Jy beam$^{-1}$km s$^{-1}$. The beam size is 4\farcs0$\times$2\farcs7. }
\label{fig:sm1_redblue}
\end{figure}

\subsection{C$^{17}$O, CH$_3$OH and H$_2$CO at SM1}

The C$^{17}$O (2--1), which is typically optically thin, can be used to probe the high-density region near SM1. The optical depth $\tau$ of C$^{17}$O in the $\rho$-oph A region was obtained from the calculation of the optical depth of C$^{18}$O in section 3.3. The $\tau$(C$^{17}$O) has maximum value of 1.5 at  5\arcsec W of SM1 and is smaller than unity in most part of the $\rho$-oph A region. The integrated intensity map of C$^{17}$O  in the SM1 field is presented in Figure \ref{fig:sm1_line}a. 
The C$^{17}$O emission extends from SM1 through VLA1623-N2, VLA1623-N1, and reaches to VLA1623. 
The overall distribution of the C$^{17}$O follows that of the continuum emission.
The C$^{17}$O peak at VLA1623 coincide well with the continuum peak.
On the other hand, the C$^{17}$O peak at SM1 appears at $\sim$3\arcsec northeast of the continuum peak.
The emission component at 15\arcsec NW of SM1 is more than 17$\sigma$, but there is no counterpart of this component in the continuum or other molecular lines. The missing flux of the C$^{17}$O was  estimated by comparing the flux value of the SMA map convolved to the JCMT beam and that observed with the JCMT. It turned out that the C$^{17}$O flux recovered by the SMA is only $\sim$3$\%$ of the single-dish flux. This indicates that most of the C$^{17}$O emission comes from the spatially extended component and very little from the envelope of SM1.

We calculated the C$^{17}$O abundance from the  continuum and C$^{17}$O data observed with the SMA by following the similar steps described in section 3.3. The  T$_{\rm ex}$ is assumed to be 27 K.   The C$^{17}$O fractional abundances are 3.6$\times$10$^{-9}$, 2.4$\times$10$^{-9}$, 2.4$\times$10$^{-9}$ and 6$\times$10$^{-9}$ at VLA1623, SM1, VLA1623-N1 and VLA1623-N2, respectively. After multiplying by the $^{18}$O/$^{17}$O relative abundance of 3.65 \citep{Penzias1981}, the derived C$^{18}$O abundances are 1.3$\times$10$^{-8}$ at VLA1623, 8.8$\times$10$^{-9}$ at SM1 and VLA1623-N1,and 2.2$\times$10$^{-8}$ at VLA1623-N2, which are significantly lower than those estimated in the previous section using the combined C$^{18}$O and continuum data. 
Because the abundance values derived here use the SMA data alone, they are the local values at the compact sources. It is likely that the C$^{18}$O abundance in the compact sources are more than one order of magnitude lower than the interstellar value, even at the positions of SM1, VLA1623-N1,and VLA1623-N2.
%Because the derived abundance are from SMA data, it only trace the local value at the compact sources; that is, the missing  C$^{17}$O flux implies that, for SM1, VLA1623-N1 and N2, most of the CO flux measured with the single-dish telescope originates form spatially extended component along the line of sight.

The CH$_3$OH and H$_2$CO show a localized emission peak at the position of SM1 (Figure \ref{fig:sm1_line}b and \ref{fig:sm1_line}c).
These line also show a localized peak toward VLA1623.
The CH$_3$OH shows the secondary peak at SM1-W.
The CH$_3$OH emission is also seen near VLA1623-N2.
The spatial distribution of the CH$_3$OH emission correlate well with that of the continuum emission, except for the missing counterpart of VLA1623-N1.
On the other hand, the spatial distribution of the H$_2$CO is different; the emission extends to the southeast and southwest of SM1.
There is no counterpart of VLA1623-N1 and VLA1623-N2 in the H$_2$CO map.
In addition, the H$_2$CO shows very bright emission at 30\arcsec N of SM1, which corresponds to the northern edge of our field of view.
This emission component is also seen in the CH$_3$OH. The location of these components corresponds to the southern edge of the $\sim$2.9 km s$^{-1}$ component seen in the single-dish maps of H$_2$CO and CH$_3$OH \citep{Bergman2011}.
%\textcolor{red}{\sout{A notable thing is that the CH$_3$OH map and the H$_2$CO map show strong E-W emission at 30" N of SM1, which are strongest components in both maps; those component also correspond to the southern edge of the CH$_3$OH at 2.9 km/s in \citet{2011A&A...527A..39B}.}}
Since CH$_3$OH and H$_2$CO are good tracers for molecular outflow\citep[e.g.][]{Bachiller2001,Gerin2015}, we searched for the signature of outflow in this region. The CH$_3$OH and H$_2$CO emission was separated into the blue-shifted part and red-shifted part (Figure \ref{fig:sm1_redblue}). 
Both blue-shifted and red-shifted components are seen in the vicinity of SM1. However, their spatial distributions do not show clear bipolarity.
\section{discussions}
\subsection{Nature of the Small condensations}
Determining the fate of small condensations such as SM1, VLA1623-N1, -N2 and SM1-W in the dense core will benefit the understanding of low-mass star forming process. The mass  of each condensation is derived from the total flux listed in Table 1, using the formula,
\begin{equation}
M_{env} = \frac{S D^2}{\kappa_{\nu}B_{\nu}(T_{\rm dust})} ,
\end{equation}
where $S$ is the total flux, $D$ is the distance to the source, $\kappa_{\nu}$ is the dust mass opacity and $B_{\nu}(T_{\rm dust})$ is the Plank function at a given temperature.
The dust mass opacity $\kappa_{\nu}$ is assumed to be 0.005 cm$^{2}$ g$^{-1}$, which is the same as in \citet{Motte1998}. The dust temperature is assumed to be 27 K \citep{Andre1993}. The masses of those condensations are in the range of 10$^{-2}$--10$^{-1}$ M$_{\odot}$. For comparison, the mass of the critical Bonnor-Ebert sphere \citep{Bonnar1956} is calculated from
\begin{equation}
M_{BE} = 2.4 \frac{kT}{mG} r_c ,
\end{equation}
where k is the Boltzmann constant, T is the gas kinetic temperature, G is the gravitational constant, m is the mean molecular mass, and $r_c$ is the critical radius. The critical radius $r_c$ is simply assumed to be the same as the semi major axis of each source. The temperature is assumed to be 27 K, which is the same as that used in the mass calculation. 
As shown in Table 1, the masses of all the condensations exceed the Bonnor-Ebert mass, which implies those condensations are gravitationally bounded.

VLA1623 and VLA1623W have already been identified as non-coeval multiple system \citep{Murillo2013a}. The recent ALMA observation by \cite{Friesen2014} also found compact structure in SM1. 
In addition, SM1 is also associated with the source detected in X-ray and radio at 5 GHz \citep{Gagne2004},
These imply that SM1 has already harbored a protostar. 
VLA1623-N1 has similar properties as SM1.
This source is identified as a compact emission source 10 by \citet{Kirk2017} in their 3 mm image observed with the ALMA. In addition, the position of the X-ray source, J162627.4-242418, which is one of the unidentified X-ray sources in \citet{Gagne2004}, coincides with that of VLA1623-N1.
Recently, \citet{Kawabe2018} studied the properties of SM1 and VLA1623-N1 based on the multi-wavelength observations from radio to X-ray, and proposed that these sources are proto brown dwarfs or in the very early phase of low-mass protostars.
On the other hand, there is no clear counterpart of VLA1623-N2 and SM1-W in the ALMA images at 3 mm \citep{Kirk2017} and 1.3 mm \citep{Kawabe2018}.
The 3 mm image of \citet{Kirk2017} shows some hint of faint condensation at the position of VLA1623-N2.
In order to search for the counterpart of VLA1623-N2, we have examined the 3 mm continuum image that was made of two datasets available from the ALMA data archive.
The resulting image (Figure \ref{fig:cont_alma} in Appendix A) has higher sensitivity and resolution than those of the 3 mm image in \citet{Kirk2017}.
However, our new 3 mm image does not show clear evidence of compact source at the position of VLA1623-N2.
The 1.3 mm image presented in \citet{Kawabe2018} also does not show the corresponding source.
In addition, there is no counterpart of this source in the X-ray \citep{Gagne2004} nor in the radio \citep{Leous1991,Gagne2004}.
Although the water maser source  is found at R.A.(J2000) = 16$^h$26$^m$27.028$^s$ and Dec. (J2000) = $-$24$^{\circ}$24\arcmin24.284\arcsec\citep{Yu1997}, which is  only 1.6\arcsec W of VLA1623-N2, it is unlikely that VLA1623-N2 harbor a protostar.
Another condensation SM1-W does not have a clear counterpart in the ALMA images of 3 mm \citep{Kirk2017}, 1.3 mm \citep{Kawabe2018}, and 0.84 mm \citep{Friesen2014} as well.
This source is not detected in the X-ray \citep{Gagne2004} nor in the radio \citep{Leous1991,Gagne2004}. 
The absence of compact condensation implies that VLA1623-N2 and SM1-W are still in the prestellar stage.
%The X-ray and the water maser emission are the good sign indicating that both condensations might harbor young stellar objects (YSOs). 
%\edit1{But recent ALMA observations\citep{Kirk2017} only reveal a compact source at VLA1623-N1; no spatially compact components are found at either VLA1623-N2 or SM1-W, suggesting there is no other clear evidence of star formation in both condensations. The 3mm continuum map observed with ALMA is shown in Appendix A. \sout{The resolution of our observation are not enough to resolve the detailed structure ($<$500 AU) in both regions. However, high resolution observations with ALMA would be able to probe those two sources}.}

\subsection{Physical condition and CO depletion}

As shown in Figure \ref{fig:temp}, the peak intensity of the C$^{18}$O is enhanced to T $\sim$40--50 K at the edges of the $\rho$-Oph A ridge, while it is lower than 30 K toward the center of the ridge.
The higher temperature in the edges is likely due to external heating by stellar UV and X-ray photons. 
%The possible nearby sources of heating are the bright 2-$\mu$m sources listed in \citet{1973ApJ...184L..53G} such as Source 1(herafter Oph S1) at $\sim$ 1.5' E of SM1N, \textcolor{red}{\sout{one of the luminous infrared sources in $\rho$ Ophiuchus cloud}}. 
One of the possible heating sources is Source 1 (herafter Oph S1) at $\sim$1.5\arcmin (12,000 au) E of SM1N.
Oph S1 having a lumnosity of ${\sim}$1600 $L_{\odot}$ \citep{Wilking2005} is known to be a binary with a B4 primary and companion with a K magnitude of 8.3 \citep{Simon1995}, and is emitting both strong non-thermal radio and X-ray emission \citep{Gagne2004}.  The UV as well as higher energy photons reach the eastern surface of the $\rho$-Oph A ridge and heat the eastern edge of the ridge.
Another heating source is the B2 star HD147889 with $\sim$4500 $L_{\odot}$ \citep{Greene1989}, which is $\sim$15\arcmin (0.5 pc) to the southwest and behind the $\rho$-Oph A ridge \citep{Liseau1999}.
This source is likely to be responsible in heating the ridge from the west.
Since the continuum emission from mid-infrared, far-infrared, to millimeter in the $\rho$-Oph A region is modeled well with the external heating of S1 and HD147889 \citep{Liseau2015}, it is natural to consider that the molecular gas in the ${\rho}$-Oph A ridge is also heated from these two stars.

On the other hand, the C$^{18}$O intensity drops significantly toward the continuum ridge; especially, it is below 15 K in the region between SM1 and SM1N.
In this region, the abundances of the C$^{18}$O and N$_2$H$^+$ show clear anti-correlation.
Similar anti-correlation between C$^{18}$O (3--2) and N$_2$H$^+$ (3--2) is also seen in \citet{Liseau2015}. 
These imply that the gas temperature in the center of the $\rho$-Oph A ridge, where the external radiation is shielded, is lower than the CO freeze-out temperature of $\sim$25 K \citep{Oberg2005}.
In the cold and dense environment, N$_2$, a parent molecule of N$_2$H$^+$, is also expected to freeze-out.
Since the desorption energy ratio of N$_2$ and CO measured in the recent laboratory experiments is $\sim$0.9 \citep{Oberg2005,Fayolle2016}, the N$_2$ freeze-out temperature is only a few degrees lower than that of CO.
The clear anti-correlation between C$^{18}$O and N$_2$H$^+$ implies that sufficient amount of N$_2$ is in the gas-phase.
This is probably because that the area with the C$^{18}$O--N$_2$H$^+$ anti-correlation is still starless \citep{Kirk2017} and the freeze-out timescales of molecules are long enough to differentiate the gas-phase abundance of CO and N$_2$.
As a result, the N$_2$H$^+$ abundance increases after CO, which is the main destructor of N$_2$H$^+$, disappears. 
%The C$^{18}$O abundance is also depleted ($< 2\times$10$^{-7}$ ) within the ridge (see Fig. \ref{fig:abundance}a). The typical canonical value of C$^{18}$O is 2 $\times$ 10$^{-7}$(reference paper ?). The cold temperature($< 30K$) would reach the CO freeze-out temperature(25 K), causing CO severely depleted from gas phase. Also, the N$_2$H$^+$ showing strong emission within the CO ridge where the C$^{18}$O is pretty low in Fig. \ref{fig:c18on2h}. The abundant N$_2H^+$ molecules can efficiently destroy CO by the reaction $N_2H^+ + CO \rightarrow N_2 + HCO^+. 
SM1 is located at the eastern edge of the C$^{18}$O depletion zone.
Although the C$^{18}$O abundance at the position of SM1 derived from the combined maps of the continuum and C$^{18}$O, 1.9$\times$10$^{-7}$, is not very low as compared to the interstellar value of 1.7--4.8$\times$10$^{-7}$, this is because of the spatially extended emission along the line of sight. 
The C$^{18}$O abundance at SM1 is a factor of 20 lower if it is derived from the SMA data. This suggests that the CO depletion is also significant in the dense gas envelope of SM1.
The C$^{18}$O abundance shows the lowest value of 4.3$\times$10$^{-8}$ toward VLA1623.
Although the C$^{18}$O line traces the rotating disk around VLA1623A, there is no C$^{18}$O emission from VLA1623B and the envelope surrounding this binary system \citep{Murillo2013a,Murillo2013}.
This implies that most of the dense gas surrounding the VLA1623A\&B binary system is colder than the CO freeze-out temperature.
Interestingly, the N$_2$H$^+$ also shows the lowest abundance at VLA1623.
This implies that the temperature of the dense gas surrounding VLA1623 is lower than the freeze-out temperature of N$_2$.
Once both CO and N$_2$ freeze-out completely, the N$_2$H$^+$ abundance does not increase when the gas is heated by the newly born star.
Since the desorption energy of N$_2$ does not differ significantly from that of CO, the N$_2$ desorption area is expected to be comparable to the CO desorption area.
In such a case, N$_2$H$^+$ is easily destructed by the CO desorbed from grains, and cannot have enough abundance to be observed.

Molecular clouds irradiated externally by nearby stars are often found in high-mass star forming regions.
For example, molecular gas in the S255-S257 system is located between two HII regions, S255 and S257, and forms an elongated ridge compressed and illuminated by two HII regions \citep{Minier2006}.
The dense gas ridge of this system harbors a cluster of YSOs with more than one hundred members \citep{Ojha2011}.
The morphology of molecular ridge and the spatial distribution of the YSOs suggest that the star formation activity in this system is induced by the compression from two HII regions.
On the other hand, dense molecular ridge irradiated by two early type stars in both sides is barely seen in the low-mass star forming regions; $\rho$-Oph A is almost a unique example. 
In the case of $\rho$-Oph A, the effect of external sources is moderate.
Molecular gas is interacting with the Photon Dominated Region (PDR) around the nearest star S1.
The H\textsc{ii} region around this star has a diameter of $<$20\arcsec \citep{Andre1988}, which is much smaller than the size of the PDR.
The curved ridge morphology observed in the C$^{18}$O (3--2) follows the outer edge of the spherical shell of the PDR around S1 traced by H$_2$, [O\textsc{i}] \citep{Larsson2017}, and [C\textsc{ii}] \citep{Mookerjea2018}, suggesting that the molecular ridge has been compressed by the PDR.
However, our C$^{18}$O (2--1) data do not show the kinematical signature of external compression. 
%\edit1{There are few similar systems among high-mass star forming regions such as S255-S257, where the molecular cloud is compressed and illuminated by two HII regions, S255 and S257 \citep{Minier2006}. \citet{Ojha2011} discovered 109 YSOs in an area of $\sim$4.9$\times$4.9 around S255-S257. The morphology of molecular material and the distribution of the YSOs both indicates the star formation activity occurred due to the collision of two swept-up bubbles. On the other hand, $\rho$-Oph A is the unique case among low-mass star forming regions. In the case of $\rho$-Oph A, the outer ridge of $\rho$-Oph A is mainly photon-dominated. The current C \textsc{ii} observations \citep{Mookerjea2018} reveals that S1 ionizes carbon and produce C \textsc{ii} region around the star, but the H \textsc{ii} region produced by S1, with a radius of $<20"$, is smaller and weaker than the C \textsc{ii} region. The detailed relation between the illuminating sources and star formation in $\rho$-oph A still needs further investigation.}

%However, at VLA1623, both C$^{18}$O and N$_2$H$^+$ show strong depletion in our map. One possibility might be due to the high dust density and the super cold temperature, probably below 15 K in those regions, which cause even N$_2$H$^+$ frozen out onto grain. However, how N$_2$H$^+$ react with other molecules below 15 K is still not clear so far \citep{2005ApJ...621L..33O}.

% S1 16 26 34.167, -24 23 28.26

\section{conclusions}

We have studied the physical and chemical conditions of the $\rho$-Oph A region using the 1.3 mm continuum, and molecular lines such as C$^{18}$O (2--1)  and N$_2$H$^+$(1--0).
The 1.3 mm continuum and C$^{18}$O (2--1) data observed with the SMA were combined with the data obtained with the IRAM 30 m telescope and JCMT, respectively, in order to fill the short-spacing information.
Our main conclusions are summarized as follows.
%We have obtained and combined the sub-millimeter observation from both high resolution interferometer such as SMA and single-dish telescope, including JCMT 30m and IRAM 15m. 
\begin{enumerate}
\item 
The combined 1.3 mm map reveals that the three major sources, VLA1623, SM1, and SM1N are embedded in the extended emission running along the north-south direction.
VLA1623 and SM1 contain the spatially compact components, while SM1N does not have a compact component.
The continuum emission around the SM1 peak is extended toward the northwest direction.
The secondary peak, SM1-W, to the west of the SM1 peak  implies the presence of small-scale clumps.
However, the location of SM1-W does not coincide with those of the clumps reported in \citet{Nakamura2012}.
\item 
In addition to VLA1623 and SM1, two compact condensations, VLA1623-N1 and VLA1623-N2, are identified in the continuum ridge connecting VLA1623 and SM1.
In addition, VLA1623W is also identified as a separate peak.
All of the small condensations identified in the $\rho$-Oph A region are gravitationally bounded.
Among the newly discovered small condensations, VLA1623-N1 has a X-ray counterpart and is likely to harbor a young stellar object.
On the other hand, VLA1623-N2 and SM1-W are starless.
%The combined maps not only have short spacing information but also maintain small synthesized beam size of 3$\sim$4\arcsec. The 1.3 mm continuum map from both 4-pointing mosaic and the newest observation reveals that SM1, in addition to the compact component, have extended structure; there are also two new found small condensation called VLA1623-N1 and VLA1623-N2 with close X-ray and water maser counterpart, respectively. Those small component including SM1, VLA1623, VLA1623W, VLA1623-N1 and VLA1623-N2 are gravitationally bound according to the comparison of their calculated mass and the mass of critical Bonner-Ebert sphere. 
\item  
The spatial distribution of the C$^{18}$O is significantly different from that of the continuum; the C$^{18}$O emission is enhanced at the eastern and western edges of the continuum ridge. The brightness temperature of the C$^{18}$O line is enhanced to 40--50 K  in the eastern and western edges.
This is consistent with the picture that the gas in the $\rho$-Oph A ridge is heated externally by the nearby B-stars Oph S1 and HD147889.
\item 
On the other hand, the gas in the inner dense ridge remains cold.
The C$^{18}$O intensity drops below 15 K in the region between SM1 and SM1N.
In this region, the C$^{18}$O abundance decreases and shows clear anti-correlation with the N$_2$H$^+$ abundance.
\item 
Both C$^{18}$O and N$_2$H$^+$ abundances decrease significantly toward the Class 0 protostar VLA1623.
This implies that most of the dense gas surrounding VLA1623 is colder than the freeze-out temperature of N$_2$, which is a few degrees lower than that of CO.
\item
The velocity structure of C$^{18}$O suggests that the $\rho$-oph A ridge consists of three components with different  systemic velocities. 
These three velocity components are known to have different chemical properties; the northern component at 3.1 km s$^{-1}$ is deficient in deuterated species, the southern component at 3.7 km s$^{-1}$ is rich in deuterated species, and the north-western component at 2.9 km s$^{-1}$ shows high abundance in methanol \citep{Bergman2011,Garay2002}.
\item
The CH$_3$OH shows the blue-shifted and red-shifted components in the vicinity of SM1.
These components are likely to trace the outflow activity of the embedded protostar in SM1.
The similar components are also seen in the H$_2$CO. 
However, the spatially extended H$_2$CO emission does not show clear bipolarity.
%The high brightness temperature feature of the eastern C$^{18}$O ridge might be related to the external heating from the bright source Oph S1. The important discovery of this paper is that the C$^{18}$O abundance is lower than the canonical value, 2$\times$10$^{-7}$, within the rho-oph A ridge; the cold temperature($< 30K$) and the anti-correlation with N$_2$H$^+$ will suggest that C$^{18}$O were frozen out or destroyed by the reaction with N$_2$H$^+$. However, both C$^{18}$O and N$_2$H$^+$ are depleted near VLA1623, which might due to super cold temperature, even N$_2$H$^+$ were frozen out.
\end{enumerate}

\acknowledgments
We thank James Di Francesco for providing the N$_2$H$^+$ data cube obtained with the IRAM 30 m telescope and the BIMA interferometer, and 1.3 mm continuum data observed with the IRAM 30 m telescope.
We thank all the SMA staff in Hawaii, Cambridge, and Taipei for their kind help during the observations. We thank the anonymous referee for the constructive suggestions.
The Submillimeter Array is a joint project between the Smithsonian Astrophysical Observatory and the Academia Sinica Institute of Astronomy and Astrophysics and is funded by the Smithsonian Institution and the Academia Sinica \citep{Ho2004}.
%The JCMT is operated by the Joint Astronomy Centre on behalf of the Science and Technology Facilities Council of the United Kingdom, the Netherlands Organisation for Pure Re- search, and the National Reserach Council of Canada.
The James Clerk Maxwell Telescope has historically been operated by the Joint Astronomy Centre on behalf of the Science and Technology Facilities Council of the United Kingdom, the National Research Council of Canada and the Netherlands Organisation for Scientific Research.
This paper makes use of the following ALMA data: ADS/JAO.ALMA\#2013.1.00187.S, ADS/JAO.ALMA\#2016.1.01468.S. ALMA is a partnership of ESO (representing its member states), NSF (USA) and NINS (Japan), together with NRC (Canada), MOST and ASIAA (Taiwan), and KASI (Republic of Korea), in cooperation with the Republic of Chile. The Joint ALMA Observatory is operated by ESO, AUI\/NRAO and NAOJ.
The National Radio Astronomy Observatory is a facility of the National Science Foundation operated under cooperative agreement by Associated Universities, Inc.
N.H. acknowledges a grant from the Ministry of Science and Technology (MoST) of Taiwan (MoST 107-2119-M-001-029)


\facilities{SMA, IRAM:30m, JCMT, ALMA}


\software{MIR (https://github.com/qi-molecules/sma-mir), MIRIAD \citep{Sault1996}, Astropy (The Astropy Collaboration 2013, 2018), Starlink \citep{Currie2014}, CASA \citep{Mcmullin2007} }


\appendix
\section{The 3 mm ALMA continuum map of the $\rho$-oph A ridge}
\begin{figure}[h]
\epsscale{0.75}
\plotone{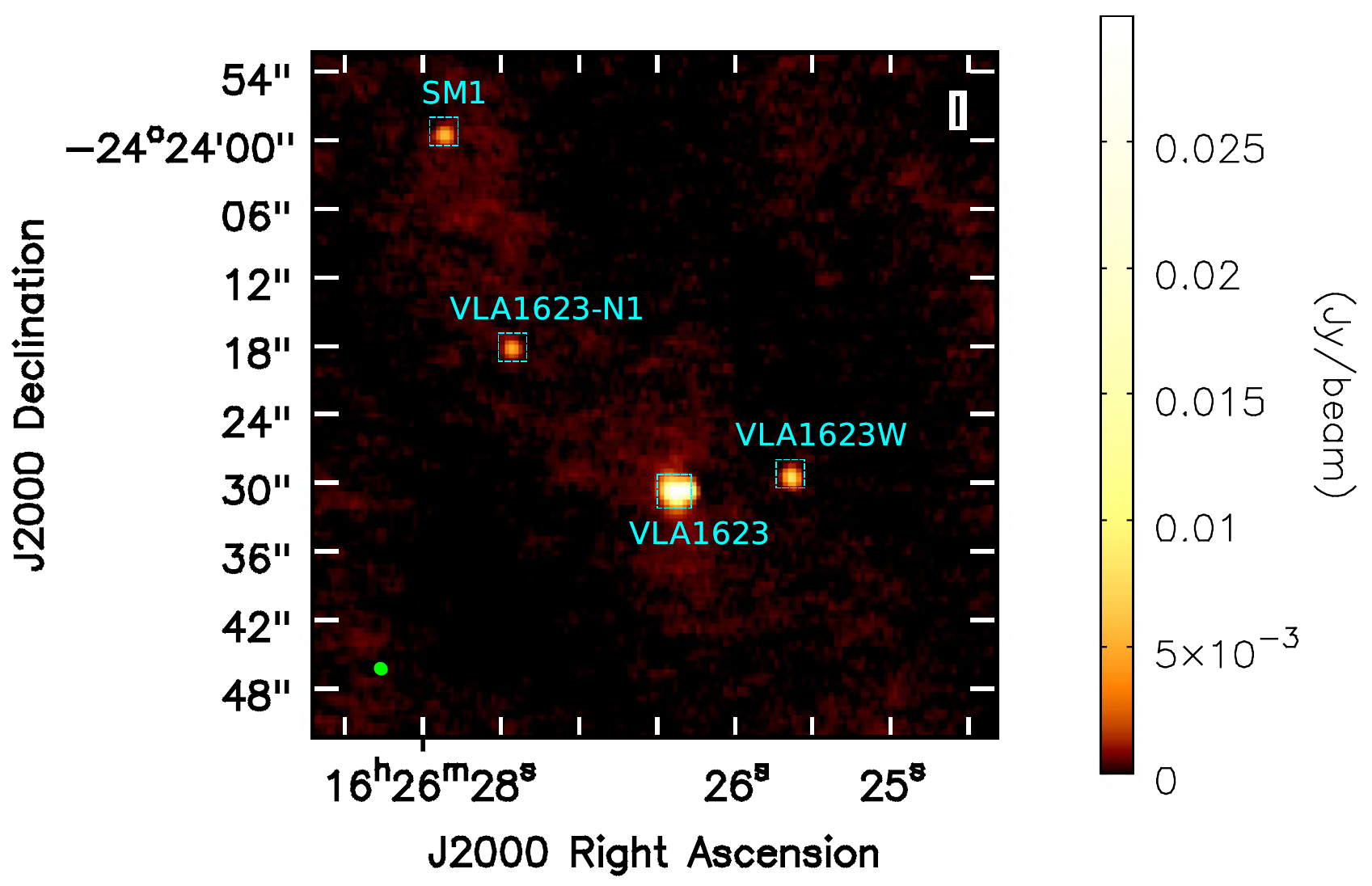}
\caption{3 mm Continuum map obtained by combining two ALMA datasets. The 1$\sigma$ rms noise level is 0.18 mJy beam$^{-1}$.  The green ellipse at bottom left corner denotes the synthesized beam size of 1\farcs1$\times$1\farcs0.}
\label{fig:cont_alma}
\end{figure}
In Figure \ref{fig:cont_alma}, we present the 3mm continuum map of the $\rho$-oph A ridge, which is made by combining ALMA data from two projects (Project code = 2013.1.00187.S and 2016.1.01468.S). The ALMA datasets were obtained from ALMA Science Archive. 
The first Band 3 observations from project 2013.1.00187.S were made during cycle 2 on January 31, 2015. The array consisted of 37 antennas with relatively compact configuration. The field center is at the position of VLA1623. The continuum data were obtained at 101, 103, and 113 GHz, each covering a bandwidth of 2GHz. The total integration time is 60.48 seconds. 
%With the angular resolution of 3\farcs8$\times$1\farcs7, the achieved continuum rms is 0.16 mJy/beam.  
The detail of the observation is also described in \citet{Kirk2017}.

The second Band 3 observations from project 2016.1.01468.S were made during cycle 4 on November 16, 2016. The array was in C40-4 configuration. 
%which gives the synthesized beam of 1\farcs1$\times$1\farcs0. 
The field is also centered at the position of VLA1623. The observation covers frequency range from 108.72 GHz to 110.94 GHz. The total integration time is 241.92 seconds
%and the achieved continuum rms is 0.44 mJy/beam.

CASA \citep{Mcmullin2007} was used to calibrate and reduce both datasets. We concatenated two calibrated datasets and used ^^ ^^ tclean" in the mfs (multi-frequency synthesis) mode to generate the combined 3 mm continuum map. 
%The natural weighting is used when producing the map. 
The robust weighting of 0.5 was adopted, providing the synthesized beam size of 1\farcs1$\times$1\farcs0.
The rms noise level of the 3 mm continuum image is 0.18 mJy beam$^{-1}$.




\bibliographystyle{aasjournal}
\bibliography{reference.bib}

\begin{thebibliography}{}
\expandafter\ifx\csname natexlab\endcsname\relax\def\natexlab#1{#1}\fi
\providecommand{\url}[1]{\href{#1}{#1}}

\bibitem[{Andre {et~al.}(1988)Andre, Montmerle, Feigelson, Stine, \&
  Klein}]{Andre1988}
Andre, P., Montmerle, T., Feigelson, E.~D., Stine, P.~C., \& Klein, K.-l. 1988,
  \apj, 335, 940

\bibitem[{Andre {et~al.}(1993)Andre, Ward-Thompson, \& Barsony}]{Andre1993}
Andre, P., Ward-Thompson, D., \& Barsony, M. 1993, \apj, 406, 122

\bibitem[{Bachiller {et~al.}(2001)Bachiller, {P{\'{e}}rez Guti{\'{e}}rrez},
  Kumar, \& Tafalla}]{Bachiller2001}
Bachiller, R., {P{\'{e}}rez Guti{\'{e}}rrez}, M., Kumar, M. S.~N., \& Tafalla,
  M. 2001, \aap, 372, 899

\bibitem[{Bergman {et~al.}(2011)Bergman, Parise, Liseau, \&
  Larsson}]{Bergman2011}
Bergman, P., Parise, B., Liseau, R., \& Larsson, B. 2011, \aap, 527, A39

\bibitem[{Bonnar(1956)}]{Bonnar1956}
Bonnar, W.~B. 1956, \mnras, 116, 351

\bibitem[{Cavanagh {et~al.}(2008)Cavanagh, Jenness, Economou, \&
  Currie}]{Cavanagh2008}
Cavanagh, B., Jenness, T., Economou, F., \& Currie, M. 2008, AN, 329, 295

\bibitem[{Chen {et~al.}(2013)Chen, Arce, Zhang, Bourke, Launhardt,
  J{\o}rgensen, Lee, Foster, Dunham, Pineda, \& Henning}]{Chen2013}
Chen, X., Arce, H.~G., Zhang, Q., {et~al.} 2013, \apj, 768, 110

\bibitem[{Currie {et~al.}(2014)Currie, Berry, \& Jenness~et al.}]{Currie2014}
Currie, M.~J., Berry, D.~S., \& Jenness~et al., T. 2014, in Astronomical Data
  Analysis Software and Systems XXIII, Vol. 485, 391

\bibitem[{{Di Francesco} {et~al.}(2004){Di Francesco}, Andre, \&
  Myers}]{DiFrancesco2004}
{Di Francesco}, J., Andre, P., \& Myers, P.~C. 2004, \apj, 617, 425

\bibitem[{Fayolle {et~al.}(2016)Fayolle, Balfe, Loomis, Bergner, Graninger,
  Rajappan, \& {\"{O}}berg}]{Fayolle2016}
Fayolle, E.~C., Balfe, J., Loomis, R., {et~al.} 2016, \apj, 816, L28

\bibitem[{Frerking {et~al.}(1982)Frerking, Langer, \& Wilson}]{Frerking1982}
Frerking, M.~A., Langer, W.~D., \& Wilson, R.~W. 1982, \apj, 262, 590

\bibitem[{Friesen {et~al.}(2014)Friesen, {Di Francesco}, Bourke, Caselli,
  J{\o}rgensen, Pineda, \& Wong}]{Friesen2014}
Friesen, R.~K., {Di Francesco}, J., Bourke, T.~L., {et~al.} 2014, \apj, 797, 27

\bibitem[{Gagne {et~al.}(2004)Gagne, Skinner, \& Daniel}]{Gagne2004}
Gagne, M., Skinner, S.~L., \& Daniel, K.~J. 2004, \apj, 613, 393

\bibitem[{Garay {et~al.}(2002)Garay, Mardones, Rodriguez, Caselli, \&
  Bourke}]{Garay2002}
Garay, G., Mardones, D., Rodriguez, L.~F., Caselli, P., \& Bourke, T.~L. 2002,
  \apj, 567, 980

\bibitem[{Gerin {et~al.}(2015)Gerin, Pety, Fuente, Cernicharo, Commer{\c{c}}on,
  \& Marcelino}]{Gerin2015}
Gerin, M., Pety, J., Fuente, A., {et~al.} 2015, \aap, 577, L2

\bibitem[{Greene \& Young(1989)}]{Greene1989}
Greene, T.~P., \& Young, E.~T. 1989, \apj, 339, 258

\bibitem[{Gurney {et~al.}(2008)Gurney, Plume, \& Johnstone}]{Gurney2008}
Gurney, M., Plume, R., \& Johnstone, D. 2008, \pasp, 120, 1193

\bibitem[{Ho {et~al.}(2004)Ho, Moran, \& Lo}]{Ho2004}
Ho, P. T.~P., Moran, J.~M., \& Lo, K.~Y. 2004, \apj, 616, L1

\bibitem[{Johnstone {et~al.}(2000)Johnstone, Wilson, Moriarty‐Schieven,
  Joncas, Smith, Gregersen, \& Fich}]{Johnstone2000}
Johnstone, D., Wilson, C.~D., Moriarty‐Schieven, G., {et~al.} 2000, \apj,
  545, 327

\bibitem[{J{\o}rgensen {et~al.}(2005)J{\o}rgensen, Sch{\"{o}}ier, \& van
  Dishoeck}]{Jorgensen2005}
J{\o}rgensen, J.~K., Sch{\"{o}}ier, F.~L., \& van Dishoeck, E.~F. 2005, \aap,
  435, 177

\bibitem[{Kawabe {et~al.}(2018)Kawabe, Hara, Nakamura, Saigo, Kamazaki,
  Shimajiri, Tomida, Takakuwa, Tsuboi, Machida, {Di Francesco}, Friesen,
  Hirano, Oasa, Tamura, Tamura, Tsukagoshi, \& Wilner}]{Kawabe2018}
Kawabe, R., Hara, C., Nakamura, F., {et~al.} 2018, arXiv:1810.00573

\bibitem[{Kirk {et~al.}(2017)Kirk, Dunham, Francesco, Johnstone, Offner,
  Sadavoy, Tobin, Arce, Bourke, Mairs, Myers, Pineda, Schnee, \&
  Shirley}]{Kirk2017}
Kirk, H., Dunham, M.~M., Francesco, J.~D., {et~al.} 2017, \apj, 838, 114

\bibitem[{Lacy {et~al.}(1994)Lacy, Knacke, Geballe, \& Tokunaga}]{Lacy1994}
Lacy, J.~H., Knacke, R., Geballe, T.~R., \& Tokunaga, A.~T. 1994, \apj, 428,
  L69

\bibitem[{Ladd {et~al.}(1998)Ladd, Fuller, \& Deane}]{Ladd1998}
Ladd, E.~F., Fuller, G.~A., \& Deane, J.~R. 1998, \apj, 495, 871

\bibitem[{Larsson \& Liseau(2017)}]{Larsson2017}
Larsson, B., \& Liseau, R. 2017, \aap, 608, A133

\bibitem[{Leous {et~al.}(1991)Leous, Feigelson, Andre, \&
  Montmerle}]{Leous1991}
Leous, J.~A., Feigelson, E.~D., Andre, P., \& Montmerle, T. 1991, \apj, 379,
  683

\bibitem[{Liseau {et~al.}(2010)Liseau, Larsson, Bergman, Pagani, Black,
  Hjalmarson, \& Justtanont}]{Liseau2010}
Liseau, R., Larsson, B., Bergman, P., {et~al.} 2010, \aap, 510, A98

\bibitem[{Liseau {et~al.}(1999)Liseau, White, Larsson, Sidher, Olofsson, Kaas,
  Nordh, Caux, Lorenzetti, Molinari, Nisini, \& Sibille}]{Liseau1999}
Liseau, R., White, G.~J., Larsson, B., {et~al.} 1999, \aap, 344, 342

\bibitem[{Liseau {et~al.}(2015)Liseau, Larsson, Lunttila, Olberg, Rydbeck,
  Bergman, Justtanont, Olofsson, \& de~Vries}]{Liseau2015}
Liseau, R., Larsson, B., Lunttila, T., {et~al.} 2015, \aap, 578, A131

\bibitem[{Loren {et~al.}(1990)Loren, Wootten, \& Wilking}]{Loren1990}
Loren, R.~B., Wootten, A., \& Wilking, B.~A. 1990, \apj, 365, 269

\bibitem[{Mangum \& Shirley(2015)}]{Mangum2015}
Mangum, J.~G., \& Shirley, Y.~L. 2015, \pasp, 127, 266

\bibitem[{Mcmullin {et~al.}(2007)Mcmullin, Waters, Schiebel, Young, \&
  Golap}]{Mcmullin2007}
Mcmullin, J.~P., Waters, B., Schiebel, D., Young, W., \& Golap, K. 2007, in
  Astronomical Data Analysis Software and Systems XVI, Vol. 376, 127

\bibitem[{Mezger {et~al.}(1992)Mezger, Sievers, Zylka, Haslam, Kreysa, \&
  Lemke}]{Mezger1992}
Mezger, P., Sievers, A., Zylka, R., {et~al.} 1992, \aap, 265, 743

\bibitem[{Minier {et~al.}(2006)Minier, Peretto, Longmore, Burton, Cesaroni,
  Goddi, Pestalozzi, \& Andr{\'{e}}}]{Minier2006}
Minier, V., Peretto, N., Longmore, S.~N., {et~al.} 2006, Proceedings of the
  International Astronomical Union, 2, 160

\bibitem[{Mookerjea {et~al.}(2018)Mookerjea, Sandell, Vacca, Chambers, \&
  G{\"{u}}sten}]{Mookerjea2018}
Mookerjea, B., Sandell, G., Vacca, W., Chambers, E., \& G{\"{u}}sten, R. 2018,
  \aap, 616, A31

\bibitem[{Motte {et~al.}(1998)Motte, Andre, \& Neri}]{Motte1998}
Motte, F., Andre, P., \& Neri, R. 1998, \aap, 336, 150

\bibitem[{Murillo \& Lai(2013)}]{Murillo2013a}
Murillo, N.~M., \& Lai, S.-P. 2013, \apj, 764, L15

\bibitem[{Murillo {et~al.}(2013)Murillo, Lai, Bruderer, Harsono, \& van
  Dishoeck}]{Murillo2013}
Murillo, N.~M., Lai, S.-P., Bruderer, S., Harsono, D., \& van Dishoeck, E.~F.
  2013, \aap, 560, A103

\bibitem[{Nakamura {et~al.}(2012)Nakamura, Takakuwa, \& Kawabe}]{Nakamura2012}
Nakamura, F., Takakuwa, S., \& Kawabe, R. 2012, \apj, 758, L25

\bibitem[{{\"{O}}berg {et~al.}(2005){\"{O}}berg, van Broekhuizen, Fraser,
  Bisschop, van Dishoeck, \& Schlemmer}]{Oberg2005}
{\"{O}}berg, K.~I., van Broekhuizen, F., Fraser, H.~J., {et~al.} 2005, \apj,
  621, L33

\bibitem[{Ojha {et~al.}(2011)Ojha, Samal, Pandey, Bhatt, Ghosh, Sharma, Tamura,
  Mohan, \& Zinchenko}]{Ojha2011}
Ojha, D.~K., Samal, M.~R., Pandey, A.~K., {et~al.} 2011, \apj, 738, 156

\bibitem[{Ortiz-Le{\'{o}}n {et~al.}(2017)Ortiz-Le{\'{o}}n, Loinard, Kounkel,
  Dzib, Mioduszewski, Rodr{\'{i}}guez, Torres, Gonz{\'{a}}lez-L{\'{o}}pezlira,
  Pech, Rivera, Hartmann, Boden, {Evans II}, Brice{\~{n}}o, Tobin, Galli, \&
  Gudehus}]{Ortiz-Leon2017}
Ortiz-Le{\'{o}}n, G.~N., Loinard, L., Kounkel, M.~A., {et~al.} 2017, \apj, 834,
  141

\bibitem[{Penzias(1981)}]{Penzias1981}
Penzias, A.~A. 1981, \apj, 249, 518

\bibitem[{Sault {et~al.}(1996)Sault, Staveley-Smith, \& Brouw}]{Sault1996}
Sault, R.~J., Staveley-Smith, L., \& Brouw, W.~N. 1996, \aaps, 120, 375

\bibitem[{Shu {et~al.}(1987)Shu, Adams, \& Lizano}]{Shu1987}
Shu, F.~H., Adams, F.~C., \& Lizano, S. 1987, \araa, 25, 23

\bibitem[{Simon {et~al.}(1995)Simon, Ghez, Leinert, Cassar, Chen, Howell,
  Jameson, Matthews, Neugebauer, \& Richichi}]{Simon1995}
Simon, M., Ghez, A.~M., Leinert, C., {et~al.} 1995, \apj, 443, 625

\bibitem[{Takakuwa(2003)}]{Takakuwa2003}
Takakuwa, S. 2003, NRO Technical Report, No.65

\bibitem[{Vogel {et~al.}(1984)Vogel, Wright, Plambeck, \& Welch}]{Vogel1984}
Vogel, S.~N., Wright, M. C.~H., Plambeck, R.~L., \& Welch, W.~J. 1984, \apj,
  283, 655

\bibitem[{Wannier(1980)}]{Wannier1980}
Wannier, P.~G. 1980, \araa, 18, 399

\bibitem[{White {et~al.}(2015)White, Drabek-Maunder, Rosolowsky, Ward-Thompson,
  Davis, Gregson, Hatchell, Etxaluze, Stickler, Buckle, Johnstone, Friesen,
  Sadavoy, Natt, Currie, Richer, Pattle, Spaans, Francesco, \&
  Hogerheijde}]{White2015}
White, G.~J., Drabek-Maunder, E., Rosolowsky, E., {et~al.} 2015, \mnras, 447,
  1996

\bibitem[{Wilking {et~al.}(2005)Wilking, Meyer, Robinson, \&
  Greene}]{Wilking2005}
Wilking, B.~A., Meyer, M.~R., Robinson, J.~G., \& Greene, T.~P. 2005, \aj, 130,
  1733

\bibitem[{Wilson {et~al.}(1999)Wilson, Avery, Fich, Johnstone, Joncas, Knee,
  Matthews, Mitchell, Moriarty-Schieven, \& Pudritz}]{Wilson1999}
Wilson, C.~D., Avery, L.~W., Fich, M., {et~al.} 1999, \apj, 513, L139

\bibitem[{Yu \& Chernin(1997)}]{Yu1997}
Yu, T., \& Chernin, L.~M. 1997, \apj, 479, L63

\end{thebibliography}



\end{document}